\documentclass[altaffilletter,aps,nofootinbib,twocolumn,prd,eqsecnum,preprintnumbers,superscriptaddress,10pt]{revtex4-1}
\pdfoutput=1
\usepackage[caption=false]{subfig}
\usepackage{graphicx}
\usepackage{amsmath}
\usepackage{amsfonts}
\usepackage{mathtools}
\usepackage{amssymb}
\usepackage{enumerate}
\usepackage{color}
\usepackage{bm}
\usepackage{mathrsfs}
\usepackage{epstopdf}
\usepackage{url}
\usepackage{footnote}
\usepackage{textcomp}
\usepackage{dsfont}
\usepackage{ulem}
\usepackage{hyperref}
\usepackage{enumerate}   
\usepackage{appendix}
\usepackage{textcomp}
\usepackage{tipa}

\makeatletter
\newcommand*{\rom}[1]{\expandafter\@slowromancap\romannumeral #1@}
\makeatother

\begin{document}

\title{Eikonal quasinormal modes and photon orbits of deformed Schwarzschild black holes
}

\author{Che-Yu Chen}
\email{b97202056@gmail.com}
\affiliation{Institute of Physics, Academia Sinica, Taipei 11529, Taiwan}

\author{Hsu-Wen Chiang}
\email{b98202036@ntu.edu.tw}
\affiliation{Leung Center for Cosmology and Particle Astrophysics, National Taiwan University, Taipei 10617, Taiwan}
\affiliation{Department of Physics and Center for Theoretical Physics, National Taiwan University, Taipei 10617, Taiwan,}

\author{Jie-Shiun Tsao}
\email{tsaojieshiun@gmail.com}
\affiliation{Department of Physics, National Taiwan Normal University, Taipei 116, Taiwan}
\affiliation{Center of Astronomy and Gravitation, National Taiwan Normal University, Taipei 116, Taiwan}

\begin{abstract}
The geometric optics approximation provides an interpretation for eikonal correspondence that, in black-hole-containing spacetimes, connects high-frequency black hole quasinormal modes with closed photon orbits around said black hole. This correspondence has been identified explicitly for Schwarzschild, Reissner-Nordstr\"om, Kerr, and Kerr-Newman black holes, the violation of which can be a potential hint toward physics beyond General Relativity. Notably, the aforementioned black hole spacetimes have sufficient symmetries such that both the geodesic equations and the master wave equations are separable. The identification of the correspondence seems to largely rely on these symmetries. One naturally asks how the eikonal correspondence would appear if the spacetime were less symmetric. For a pioneering work in this direction, we consider in this paper a deformed Schwarzschild spacetime retaining only axisymmetry and stationarity. We show that up to the first order of spacetime deformations the eikonal correspondence manifests through the definition of the \textit{averaged} radius of trapped photon orbits along their one period. This averaged radius overlaps the potential peak in the master wave equation, which can be defined up to the first order of spacetime deformations, allowing the explicit identification of the eikonal correspondence. 
\end{abstract}

\maketitle

\section{Introduction}

The recent direct detection of gravitational waves emitted from the mergers of binary black holes is a tremendous achievement in modern physics \cite{LIGOScientific:2016aoc,LIGOScientific:2018mvr,LIGOScientific:2020ibl,LIGOScientific:2021djp}. It allows us to probe strong gravity regimes, e.g., the vicinity of black holes, through entirely different ways than traditional electromagnetic observations \cite{Barack:2018yly}. In particular, the gravitational wave signals at the post-merger phase, i.e., the ringdown signals, can be a promising tool to explore black holes, or even to test the underlying gravitational theories \cite{Berti:2018vdi}. During the post-merger phase, the two objects in the binary have already merged and formed typically a final black hole. Before the newly formed black hole settles into its stationary configuration, the distortions in its shape relax under a certain characteristic pattern, a superposition of sinusoidal oscillations with exponentially decaying amplitudes known as the quasinormal modes (QNMs) \cite{Kokkotas:1999bd,Berti:2009kk,Konoplya:2011qq}. Black hole QNMs have complex-valued frequencies, with the real parts describing the oscillations and the imaginary parts corresponding to the decay. One important feature of the black hole QNM spectrum that cements it as a powerful tool to test gravitational theories is that in General Relativity (GR) QNM spectra satisfy the black hole no-hair theorem. More explicitly, the spectra only depend on the mass, charge, and spin of the black hole, no matter what mechanism triggers the distortions in the first place. Various works have been devoted to the testing of GR or Kerr hypothesis through black hole QNMs \cite{Dreyer:2003bv,Blazquez-Salcedo:2017txk,Glampedakis:2017dvb,Tattersall:2018nve,Cardoso:2019mqo,Chen:2019iuo,Isi:2019aib,McManus:2019ulj,Maselli:2019mjd,Cabero:2019zyt,Tattersall:2019nmh,Bouhmadi-Lopez:2020oia,Chen:2020evr,Isi:2020tac,Bamber:2021knr,Chen:2021cts,Pierini:2021jxd,Ikeda:2021uvc,Ghosh:2021mrv,Cano:2021myl,Momennia:2022tug,Arbey:2021jif}, although still no evidence of physics beyond GR has been found \cite{LIGOScientific:2020tif,LIGOScientific:2021sio}.

Another equally important achievement in modern physics is the first observational image of M87*, the supermassive black hole at the center of the M87 galaxy, released from the Event Horizon Telescope Collaboration \cite{EventHorizonTelescope:2019dse}. The released image can already resolve a bright ring encircling a dark spot, which indicates a spacetime region with a tremendously strong gravitational field. The black hole \textit{shadow} image is completely due to the gravitational lensing effects near the black hole and thus can also be a potential tool to probe black hole spacetimes \cite{Cunha:2018acu}. In particular, on the image plane, somewhere near the bright ring, there is a critical curve around the dark spot, and the observation of this curve, theoretically speaking, requires perfect resolution. The critical curve is essentially the impact parameter of the photon region around the black hole, in which photons can undergo spherical motions due to strong lensing effects \cite{Perlick:2021aok}. The critical curves in black hole shadow images only carry information about the black hole geometry and do not depend on the details of the astrophysical environment. Therefore, the critical curves in shadow images have been widely investigated for various black hole models \cite{Johannsen:2013vgc,Li:2013jra,Younsi:2016azx,Wang:2017hjl,Tsukamoto:2017fxq,Abdikamalov:2019ztb,Shaikh:2019fpu,Bambi:2019tjh,Vagnozzi:2019apd,Kumar:2019pjp,Liu:2020ola,EslamPanah:2020hoj,Khodadi:2020jij,Jusufi:2020odz,EventHorizonTelescope:2020qrl,Khodadi:2020gns,Hu:2020usx,Brahma:2020eos,Lima:2021las,Konoplya:2021slg,EventHorizonTelescope:2021dqv,Lara:2021zth,Cimdiker:2021cpz,Meng:2022kjs}. Recently, some works have been devoted to testing spacetime symmetries or some putative principles using black hole images by looking for unique features on the critical curves \cite{Chen:2020aix,Eichhorn:2021etc,Eichhorn:2021iwq,Lin:2022ksb,Eichhorn:2022oma}.   

The above aspects of black holes, i.e., the black hole QNMs and the shadow images, although seemingly different, are intimately related. In Ref.~\cite{Ferrari:1984zz}, Ferrari and Mashhoon identified an analytic relation between the high-frequency QNMs, or eikonal QNMs, and the spherical photon orbits for several kinds of black holes, such as the Schwarzschild, Reissner-Nordstr\"om, and slowly rotating black holes. The validity of this relation stems from the geometric optics approximation of waves propagating around a black hole \cite{Hod:2009td}, which can be described as scattering processes \cite{Gaddam:2020mwe}. In Ref.~\cite{Cardoso:2008bp}, the eikonal correspondence was generalized to stationary, spherically symmetric, and asymptotically flat spacetimes with arbitrary dimensions. The real parts of QNM frequencies can be formally related to the orbital frequency on the spherical photon orbits, and the imaginary parts can be identified as the Lyapunov exponent on the orbits \cite{Dolan:2009nk}. Investigating this correspondence in Kerr spacetimes with an arbitrary spin is not a trivial task. In Ref.~\cite{Dolan:2010wr}, the eikonal correspondence of Kerr black holes has been identified for equatorial orbits ($l=|m|$) and polar orbits ($m=0$), where $l$ is the multipole number and $m$ is the azimuthal number of the modes. Later on, taking advantage of the separability of geodesic equations and master wave equations of Kerr spacetimes, the eikonal correspondence of Kerr black holes for arbitrary spins and modes has been fully explored \cite{Yang:2012he} (see also Ref.~\cite{Li:2021zct} for a more recent investigation on Kerr-Newman spacetimes). In addition, since the critical curves in the shadow images of black holes are determined by the impact parameter of spherical photon orbits around black holes, it is not surprising that the eikonal QNMs can also be related to the shadow cast by the black hole. The correspondence between these two seemingly different contexts has recently gained several interests \cite{Stefanov:2010xz,Jusufi:2019ltj,Jusufi:2020dhz,Cuadros-Melgar:2020kqn,Yang:2021zqy,Li:2021mnx} due to its possible future astrophysical implications \cite{Yang:2021zqy,Zhang:2021ygh}. In particular, the violation of eikonal correspondence can be a smoking gun of physics beyond GR \cite{Konoplya:2017wot,Glampedakis:2019dqh,Chen:2019dip,Silva:2019scu,Chen:2021cts,Bryant:2021xdh,Moura:2021eln}. More explicitly, this correspondence can be violated when gravitons effectively pick up some non-minimal couplings with other degrees of freedom \cite{Konoplya:2017wot,Moura:2021eln,Bryant:2021xdh,Glampedakis:2019dqh,Silva:2019scu}, or when non-minimal matter couplings are directly imposed in the theory \cite{Chen:2021cts,Chen:2019dip}. These two scenarios are clear features of non-GR physics and can be tested via the eikonal correspondence.

It should be emphasized that the explicit identification of the eikonal correspondence discussed previously relies on the symmetry of the black hole spacetime under consideration. For example, for static and spherically symmetric black holes, the master wave equation can be simplified via suitable field redefinitions and separations of variables. The QNMs for these cases are typically described by a Schr\"odinger-like equation, in which the effective potential directly determines the QNM spectrum after imposing proper boundary conditions. Because of the spherical symmetry of the spacetime, the spherical photon orbits have a unique radius, forming a shell around the black hole, i.e., the photon sphere. In the eikonal limit, the peak of the effective potential in the master wave equation is precisely at the photon sphere, naturally leading to the eikonal correspondence. For Kerr and Kerr-Newman black holes, the general identification of the eikonal correspondence performed in Refs.~\cite{Yang:2012he,Li:2021zct} takes advantage of the separability of geodesic equations and the master wave equations in the eikonal limit. The separability of geodesic equations is tightly related to the existence of the Carter constant, which corresponds to an additional hidden symmetry of the spacetime. 

Based on these results, one naturally asks how the eikonal correspondence would manifest in a black hole spacetime with fewer symmetries where neither the geodesic equation nor the wave equation is separable? In this paper, as a first step toward addressing this question, we will demonstrate how to identify the eikonal correspondence in the scenario where a Schwarzschild spacetime picks up a general axisymmetric stationary deformation.

The challenges of this work are twofold. First, due to the arbitrary spacetime deformations, the master wave equation cannot be recast as a Schr\"odinger-like form. However, it has been shown recently in Ref.~\cite{Cano:2020cao} that a Schr\"odinger-like expression for the master equation could be attainable if the deformations on the original separable equations are small. This method allows us to compute the shift to the QNM spectra induced by deformations to the Schwarzschild spacetime. Second, the radial and the polar sectors of geodesic equations, in general, cannot be decoupled. The identification of constant radial motions is only possible for very limited orbits (see also Ref.~\cite{Glampedakis:2018blj} for relevant discussions). In this work, we will show that if the orbits on the \textit{deformed} photon sphere are periodic and form limit cycles, one can define an averaged radius for these orbits and identify them as the peak of the effective potential in the master wave equation, leading to the eikonal correspondence. 

This paper is outlined as follows. In sec.~\ref{sec.geoapp}, we first briefly review the geometric optics approximation in generic curved spacetimes, then take a general spherically symmetric spacetime as an example to demonstrate its eikonal correspondence. In sec.~\ref{sec.spacetime}, we introduce the deformed Schwarzschild spacetime considered throughout this paper. In sec.~\ref{sec.kg}, we approximate the massless Klein-Gordon equation and recast the master wave equation as a Schr\"odinger-like form by the method proposed in Ref.~\cite{Cano:2020cao}. Sec.~\ref{sec.eikonal} is devoted to obtaining the eikonal expressions of the master wave equation. Then, in sec.~\ref{sec.geodecorr}, we investigate the photon geodesic equations in this spacetime and establish the eikonal correspondence, starting with the simplest circular photon orbits, then the polar orbits, and eventually generic orbits with arbitrary inclinations. The identification of the eikonal correspondence for some specific metrics will be given in sec.~\ref{sec.example} to support our results. We finally conclude in sec.~\ref{sec.con}. We would like to mention that throughout this paper the comma and $\partial$ represent partial derivatives interchangeably for convenience, while primes and dots represent derivatives with respect to the radial coordinate $r$ and the affine parameter $\lambda$, respectively.

\section{Geometric optics approximation}\label{sec.geoapp}
In this section, we will first briefly review the geometric optics approximation (sometimes called the eikonal approximation) in generic curved spacetimes. It will be shown that the equations of motion for test fields in the eikonal limits are identical to those of freely moving photons. This property holds for both electromagnetic fields and scalar fields, and is expected in general for various types of fields.

The geometric optics approximation essentially requires the wavelength scale of the test field to be much smaller than any other length scale in the system. In this limit, the equations of motion for most test fields can be expressed in the form \cite{MTW:Gravitation}
\begin{equation}
    \nabla^\alpha\nabla_\alpha\bold{A}=0\,,\label{eikonalappgeneral}
\end{equation}
where $\nabla_\alpha$ denotes the covariant derivative defined in the spacetime. The field $\bold{A}$ represents the test field. In the eikonal approximation, the left-hand side of Eq.~\eqref{eikonalappgeneral} would typically dominate over other terms that could appear in the system (e.g., field mass, spacetime curvature, and field spin). Therefore, the field $\bold{A}$ described in Eq.~\eqref{eikonalappgeneral} is not necessarily a scalar field. It can also be Maxwell gauge fields or fields with different spins. In the eikonal approximation, their evolution equations can be expressed as Eq.~\eqref{eikonalappgeneral}.

The test field function can be further decomposed as
\begin{equation}
    \bold{A}=|\bold{a}|e^{iS}\,,
\end{equation}
where the amplitude $|\bold{a}|$ and the phase $S$ are functions of spacetime coordinates. The phase $S$ varies on the scale of field wavelength, while the amplitude is slowly changing.
 
Then, one defines the wave vector $k_\mu=\partial_\mu S$. Eq.~\eqref{eikonalappgeneral} at the leading order and the next-to-leading order in the eikonal limit can be written as
\begin{align}
    g^{\mu\nu}k_\mu k_\nu=k^\mu k_\mu=0\,,\label{eikonaleading}\\
    2k^\mu\partial_\mu\ln{|\bold{a}|}+\nabla_\mu k^\mu=0\,,\label{eikonalsl}
\end{align}
respectively. The leading order equation \eqref{eikonaleading} is equivalent to the geodesic equations for null rays:
\begin{equation}
    k^\mu\nabla_\mu k_\nu=k^\mu\nabla_\mu\nabla_\nu S=k^\mu\nabla_\nu\nabla_\mu S=k^\mu\nabla_\nu k_\mu=0\,.
\end{equation}
Therefore, in the eikonal limit, the wave solution of Eq.~\eqref{eikonalappgeneral} naturally corresponds to the light rays propagating in the spacetime. 

It is well known that black hole QNMs can be treated naively as test fields scattered around black holes. Therefore, based on the geometric optics approximation, the eikonal modes can be interpreted as wave packets localized near the spherical photon orbits. As mentioned in the Introduction, this correspondence between eikonal black hole QNMs and photon orbits has been well understood for Schwarzschild black holes and several black hole spacetimes beyond Schwarzschild that preserve spherical symmetry. The eikonal correspondence for rotating black holes is much more non-trivial and was just fully uncovered in recent years e.g. see \cite{Yang:2012he} and \cite{Li:2021zct} for the investigation on Kerr and Kerr-Newman black holes, respectively.

\subsection{Eikonal correspondence: Spherically symmetric black holes}\label{subsec.sss}
In this subsection, we briefly review how the correspondence between eikonal black hole QNMs and the photon orbits manifests for a spherically symmetric black hole. The discussion strictly follows that in Ref.~\cite{Glampedakis:2019dqh}.

We first consider the single wave equation, which generally describes a test field $\psi$ propagating in a spherically symmetric spacetime,
\begin{equation}
    \psi_{,yy}+\left(\omega^2-U\right)\psi=0\,,\label{generalwaveeq}
\end{equation}
where $\omega$ is the frequency and $U=U(y)$ is the effective potential that depends only on a radial coordinate $y$. The prime denotes partial derivatives. In the following discussion, we will assume that the potential is real valued and has only a single peak. Also, the effective potential is assumed to vanish near two asymptotic regions $y\rightarrow\pm\infty$. As mentioned before, the wave function can be parametrized as
\begin{equation}
    \psi(y)=|\bold{a}(y)|e^{iS(y)/\bar\epsilon}\,,\label{wfnwkb}
\end{equation}
where $\bar\epsilon$ tracks the order of the eikonal approximation.
By inserting the wave function \eqref{wfnwkb} into the wave equation \eqref{generalwaveeq}, we get
\begin{align}
    |\bold{a}|_{,yy}+\frac{i}{\bar\epsilon}\left(2S_{,y}|\bold{a}|_{,y}+|\bold{a}|S_{,yy}\right)&-\frac{1}{\bar\epsilon^2}|\bold{a}|S_{,y}^2\nonumber\\
    +\left(\omega^2-U\right)|\bold{a}|&=0\,.\label{wkbgeneraleq}
    \end{align}
In the leading order of the eikonal approximation, the above equation reads
\begin{equation}
    -\frac{1}{\bar\epsilon^2}S_{,y}^2+\omega^2-U=0\,.\label{wkbzeroorder}
\end{equation}
Although the frequency $\omega$ and the potential $U$ do not explicitly depend on $\bar\epsilon$, they can be expressed in terms of eikonal expansion. Only the leading order terms of them are considered in Eq.~\eqref{wkbzeroorder}. After taking the derivative with respect to $y$, we have
\begin{equation}
    \frac{2}{\bar\epsilon^2}S_{,y}S_{,yy}+U_{,y}=0\,.\label{wkbde}
\end{equation}
One can see from Eq.~\eqref{wkbzeroorder} that the leading eikonal order of $\omega^2$ is real, meaning that $\omega$ is either real or purely imaginary at the leading eikonal order. We shall focus on the scenario where $\omega$ is real in the leading eikonal order, because this is typically the case for, e.g., Schwarzschild black holes.

Because the boundary conditions for black hole QNMs are $S(y\rightarrow\pm\infty)=\pm\omega y$, there must exist a location $y_m$ where the phase function $S$ takes its minimum value, i.e., $(S_{,y})_m=0$, where the subscript $m$ means that the quantity is evaluated at $y_m$. According to Eq.~\eqref{wkbde}, this location $y_m$ is precisely the peak of the potential $U$. The real part of QNM frequencies can therefore be obtained by evaluating Eq.~\eqref{wkbzeroorder} at the potential peak,
\begin{equation}
    \omega^{(0)}_R=\sqrt{U_m}\,,\label{eikonalreal}
\end{equation}
where the superscript $(0)$ indicates the leading eikonal order of the subject, in this case, $\omega$.

Let us now consider the imaginary part of QNM frequencies, which requires the calculations at the next-to-leading eikonal order. Taking the imaginary part of the next-to-leading eikonal order of Eq.~\eqref{wkbgeneraleq}, we have
\begin{equation}
    \frac{1}{\bar\epsilon}\left(2S_{,y}|\bold{a}|_{,y}+|\bold{a}|S_{,yy}\right)+2\omega^{(0)}_R\omega^{(1)}_I|\bold{a}|=0\,,
\end{equation}
where the superscript $(1)$ indicates the next-to-leading contribution. Evaluating this equation at the potential peak $y_m$, we get
\begin{equation}
    \frac{1}{\bar\epsilon}\left(S_{,yy}\right)_{m}+2\omega^{(0)}_R\omega^{(1)}_I=0\,.
\end{equation}
The second derivative of the phase function can be obtained by taking the Taylor expansion of Eq.~\eqref{wkbzeroorder} near $y_m$. We have
\begin{equation}
    \frac{1}{\bar\epsilon}\left(S_{,yy}\right)_{m}=\sqrt{\frac{|U_{,yy}|_{m}}{2}}\,,
\end{equation}
and in conjunction, the imaginary part of QNM frequencies
\begin{equation}
    \omega^{(1)}_I=-\frac{1}{2}\sqrt{\frac{|U_{,yy}|_m}{2U_m}}\,.\label{eikonaimaginary}
\end{equation}
It should be noted that Eqs.~\eqref{eikonalreal} and \eqref{eikonaimaginary} can also be obtained using the Wentzel-Kramers-Brillouin (WKB) approach, which is a powerful semi-analytic method to calculate QNM frequencies \cite{Mashhoon:1982im,Schutz:1985km,Iyer:1986np}. The geometric optics approximation is perfectly within the range of validity of the WKB method.

According to Eqs.~\eqref{eikonalreal} and \eqref{eikonaimaginary}, one can determine the eikonal QNM frequency by the value and the second derivative of the potential at the potential peak. The well-known eikonal correspondence between black hole QNMs and the photon sphere around black holes is directly related to the fact that the potential peak $y_m$ in the eikonal limit is precisely at the photon sphere \cite{Cardoso:2008bp}. For example, the effective potential for the Schwarzschild black hole can be written as
\begin{equation}
    U(y(r))=\left(1-\frac{2M}{r}\right)\frac{l(l+1)}{r^2}+O(l^0)\,.\label{potentialSch}
\end{equation}
 In the eikonal limit, we have $\bar\epsilon l=O(1)$, so the first term on the right-hand side of Eq.~\eqref{potentialSch} dominates the remaining terms. The peak of the potential \eqref{potentialSch} is on $r=3M$, which is precisely at the photon sphere $r_{ph}$ around the Schwarzschild black hole. The right-hand side of Eqs.~\eqref{eikonalreal} and \eqref{eikonaimaginary} can be related to the orbital frequency $\Omega_{ph}$ and the Lyapunov exponent $\gamma_{ph}$ on the photon sphere, respectively \cite{Cardoso:2008bp},
 \begin{equation}
     \omega^{(0)}_R=l\Omega_{ph}\,,\quad\omega^{(1)}_I=-\frac{1}{2}|\gamma_{ph}|\,,\label{eikonalfrelya}
 \end{equation}
 where
 \begin{equation}
     \Omega_{ph}=\frac{\sqrt{f(r_{ph})}}{r_{ph}}\,,\quad \gamma_{ph}^2=-\frac{1}{2}\left\{r^2f(r)\left[\frac{f(r)}{r^2}\right]_{,rr}\right\}_{ph}\,,
 \end{equation}
 and $f(r)\equiv1-2M/r$. This correspondence does not just hold for Schwarzschild black holes. It also holds for charged black holes and even for several black hole spacetimes in theories beyond GR, although some exceptions have been found in the literature \cite{Konoplya:2017wot,Glampedakis:2019dqh,Chen:2019dip,Silva:2019scu,Chen:2021cts,Bryant:2021xdh}. 

Having shown that the waves naturally behave like photons in the eikonal limit and that this property manifests naturally as the correspondence between QNMs and photon spheres for spherically symmetric black holes, one may then ask how this correspondence appears if the black hole is less symmetric. For Kerr black holes, one even has to rely on the separability of both the geodesic equations and the wave equations to identify the correspondence \cite{Yang:2012he}. What if the black hole is deformed such that the geodesic equations and the wave equations no longer separate? Can one still define an effective potential and its peak properly? If so, would the peak correspond to some generalized spherical photon orbits? We will investigate these issues in the rest of this paper.

\section{Deformed Schwarzschild spacetime}\label{sec.spacetime}
We consider a deformed Schwarzschild spacetime whose deformations satisfy the following assumptions: (i) The deformations are small, i.e., they can be parametrized by a small dimensionless parameter $\epsilon$, (ii) the deformed Schwarzschild spacetime is stationary and axisymmetric, and (iii) no frame-dragging deformations appear. The frame-dragging deformations would induce frequency shifts to QNMs, similar to the eikonal correspondence investigated in Ref.~\cite{Ferrari:1984zz}, where Schwarzschild spacetimes are extended to slowly rotating ones. Therefore, we will not consider frame-dragging deformations in this paper.

We express the deformed Schwarzschild metric using the standard Schwarzschild coordinates $(t,r,\theta,\varphi)$,
\begin{align}
g_{tt}&=-\left(1-\frac{2M}{r}\right)\left(1+\epsilon A_j(r)\cos^j{\theta}\right)\,,\nonumber\\
g_{rr}&=\left(1-\frac{2M}{r}\right)^{-1}\left(1+\epsilon B_j(r)\cos^j{\theta}\right)\,,\nonumber\\
g_{\theta\theta}&=r^2\left(1+\epsilon C_j(r)\cos^j{\theta}\right)\,,\nonumber\\
g_{\varphi\varphi}&=r^2\sin^2\theta\left(1+\epsilon D_j(r)\cos^j{\theta}\right)\,,\nonumber\\
g_{tr}&=\epsilon a_j(r)\cos^j{\theta}\,,\qquad g_{t\theta}=\epsilon b_j(r)\cos^j{\theta}\,,\nonumber\\
g_{r\theta}&=\epsilon c_j(r)\cos^j{\theta}\,,\qquad g_{r\varphi}=\epsilon d_j(r)\cos^j{\theta}\,,\nonumber\\
g_{\theta\varphi}&=\epsilon e_j(r)\cos^j{\theta}\,,\label{parametric}
\end{align}
where $M$ is the black hole mass and the dummy index $j$ stands for summations running upward from $j=0$. The deformed spacetime remains stationary and axisymmetric, i.e., the metric does not depend explicitly on $\{t,\varphi\}$. The metric deformations in each component of the metric are expanded as a series of $\cos\theta$, with each term in the series weighted by a radial function. More specifically, the radial functions $A_j(r)$, $B_j(r)$, $C_j(r)$, and $D_j(r)$ appear in the diagonal components of the metric, while $a_j(r)$, $b_j(r)$, $c_j(r)$, $d_j(r)$, and $e_j(r)$ appear in the off-diagonal components. All these radial functions, roughly speaking, have to vanish at $r\rightarrow\infty$ in order to satisfy the asymptotic flatness condition. However, they can be $O(1)$ near the event horizon. The parameter $|\epsilon|\ll1$ is dimensionless and its smallness implies that the deformations are small. In this paper, we will only consider expansions up to the first order of $\epsilon$. Finally, we exclude the frame-dragging terms, i.e., the $g_{t\varphi}$ component, in the deformations, as we mentioned at the beginning of this section.

The metric \eqref{parametric} contains several off-diagonal components. One may wonder whether the metric can be diagonalized via some coordinate transformations. Here, we will quickly show that these off-diagonal components render the spacetime non-circular and cannot be removed by coordinate transformations. For the discussion on non-circular spacetimes and their observational implications, we refer the readers to Refs.~\cite{Anson:2020trg,BenAchour:2020fgy,Minamitsuji:2020jvf,Delaporte:2022acp}.

Following the definition in Ref.~\cite{Frolov:1998wf} (sec. 6.3.4), an axisymmetric and stationary metric satisfies the circularity condition if there exist coordinate transformations that render the metric coefficients $g_{tr}$, $g_{t\theta}$, $g_{r\varphi}$, $g_{\theta\varphi}$ zero. We follow the calculations of \cite{Anson:2020trg} by first defining the 1-forms associated with the two Killing vectors:
\begin{equation}
k=g_{t\mu}dx^\mu\,,\quad \eta=g_{\varphi\mu}dx^\mu\,.
\end{equation}
The circularity of the metric is identical to the following integrability conditions:
\begin{equation}
k\wedge\eta\wedge dk=k\wedge\eta\wedge d\eta=0\,.
\end{equation}
We find that
\begin{align}
k\wedge\eta&\wedge dk=g_{tt}g_{\varphi\varphi}\left(\frac{\partial g_{t\theta}}{\partial r}-\frac{\partial g_{tr}}{\partial\theta}\right)dt\wedge dr\wedge d\theta\wedge d\varphi\nonumber\\
&+g_{\varphi\varphi}\left(g_{tr}\frac{\partial g_{tt}}{\partial\theta}-g_{t\theta}\frac{\partial g_{tt}}{\partial r}\right)dt\wedge dr\wedge d\theta\wedge d\varphi\,,
\end{align}
and
\begin{align}
k\wedge\eta&\wedge d\eta=g_{tt}g_{\varphi\varphi}\left(\frac{\partial g_{\theta\varphi}}{\partial r}-\frac{\partial g_{r\varphi}}{\partial\theta}\right)dt\wedge dr\wedge d\theta\wedge d\varphi\nonumber\\
&+g_{tt}\left(g_{r\varphi}\frac{\partial g_{\varphi\varphi}}{\partial\theta}-g_{\theta\varphi}\frac{\partial g_{\varphi\varphi}}{\partial r}\right)dt\wedge dr\wedge d\theta\wedge d\varphi\,.
\end{align}
Up to the first order in $\epsilon$, the above equations can be expressed as
\begin{align}
k\wedge\eta&\wedge dk\approx \epsilon r^2\sin^2\theta f'b_j\cos^j\theta\nonumber\\&-\epsilon r^2f\sin^2\theta\left[a_j\sin\theta\frac{d\left(\cos^j\theta\right)}{d\cos\theta}+b_j'\cos^j\theta\right]\,,\label{circularirty1}\\
k\wedge\eta&\wedge d\eta\nonumber\\\approx&-2\epsilon r^2f\cos\theta\sin\theta d_j\cos^j\theta+2\epsilon fr\sin^2\theta e_j\cos^j\theta\nonumber\\
&-\epsilon r^2f\sin^2\theta d_j\sin\theta\frac{d\left(\cos^j\theta\right)}{d\cos\theta}-\epsilon r^2f\sin^2\theta e_j'\cos^j\theta\,,\label{circularirty2}
\end{align}
where $f(r)=1-2M/r$ and primes denote derivatives with respect to $r$. Eqs.~\eqref{circularirty1} and \eqref{circularirty2} are generically not zero. Therefore, the deformed Schwarzschild metric \eqref{parametric} is not circular. The off-diagonal components could have physical consequences.

\section{Klein-Gordon equation}\label{sec.kg}
In this paper, we consider massless scalar waves propagating in the deformed Schwarzschild spacetime \eqref{parametric}. The consideration of different kinds of fields (vectors, massive scalars, etc.) does not change the results because of the geometric optics approximation. The wave equation is governed by the massless Klein-Gordon equation
\begin{equation}
\Box\psi=0\,.
\end{equation}
Because there are two Killing vectors $\partial_t$ and $\partial_\varphi$, the wave function as well as the wave equation can be decomposed as
\begin{equation}
\Box\psi=\int_{-\infty}^\infty d\omega \sum_{m=-\infty}^\infty e^{i(m\varphi-\omega t)}\mathcal{D}_{m,\omega}^2\psi_{m,\omega}(r,\theta)\,,
\end{equation}
such that each Fourier mode of the wave function satisfies
\begin{equation}
\mathcal{D}_{m,\omega}^2\psi_{m,\omega}=0\,.
\end{equation}
Here, $m$ and $\omega$ represent the azimuthal number and the mode frequency, respectively.

In the original Schwarzschild spacetime, the Klein-Gordon equation is separable. More explicitly, using Legendre functions as angular bases, the Klein-Gordon equation can be separated into a radial equation and an angular equation. The radial equation can be further recast into a Schr\"odinger-like form with an effective potential{\footnote{For the Schwarzschild metric, it corresponds to the Regge-Wheeler equation with spin $s=0$.}}. However, once including general deformations, the deformed Schwarzschild spacetime \eqref{parametric} does not allow for separable solutions to the Klein-Gordon equation due to the generic $\{r,\theta\}$ dependence in the operator. Recently, a projection method has been proposed \cite{Cano:2020cao} to deal with an ``almost" separable system deformed from a separable one. The non-separability of the system is solely contributed by small spacetime deformations. In such cases, one can obtain the $\epsilon$-order correction terms on top of the zeroth order radial equation, and reshuffle them into a modified radial equation that encodes the QNM frequency shifts due to metric deformations up to the first order. The method has been applied in Ref.~\cite{Cardoso:2021qqu} to consider the QNMs of tidally deformed spacetimes.

Up to the first order of $\epsilon$, the operator can be written as
\begin{equation}
\mathcal{D}_{m,\omega}^2=\mathcal{D}_{(0)m,\omega}^2+\epsilon\mathcal{D}_{(1)m,\omega}^2\,,\label{D01}
\end{equation}
where the zeroth order operator reads
\begin{align}
\mathcal{D}^2_{(0)m,\omega}=&-\left(\omega^2-\frac{m^2f(r)}{r^2\sin^2{\theta}}\right)-\frac{f(r)}{r^2}\partial_{r}\left(r^2f(r)\partial_{r}\right)\nonumber\\&-\frac{f(r)}{r^2\sin{\theta}}\partial_{\theta}\left(\sin{\theta}\partial_{\theta}\right)\,.\label{KGoperator1}
\end{align}
The first order operator is
\begin{widetext}
\begin{align}
\mathcal{D}_{(1)m,\omega}^2&=\frac{m^2f(r)}{r^2\sin^2\theta}\left(A_j-D_j\right)\cos^j\theta-\frac{f(r)}{r^2}\left(A_i-B_j\right)\cos^j\theta\left[\partial_r\left(r^2f\partial_r\right)\right]-\frac{f^2}{2}\left(A_j'-B_j'+C_j'+D_j'\right)\cos^j\theta\partial_{r}\nonumber\\&-\frac{f}{r^2}\left(A_j-C_j\right)\cos^j\theta\left(\cot\theta\partial_\theta+\partial_\theta^2\right)-\frac{f}{2r^2}\left[\left(A_j+B_j-C_j+D_j\right)\partial_\theta\cos^j\theta\right]\partial_\theta-\frac{2i\omega f}{r}a_j\cos^j\theta\left(r\partial_r+1\right)\nonumber\\&-i\omega f\partial_ra_j\cos^j\theta-\frac{2i\omega}{r^2}b_j\cos^j\theta\partial_\theta-\frac{i\omega}{r^2\sin\theta}b_j\partial_\theta\left(\cos^j\theta\sin\theta\right)-\frac{imf}{r^2\sin^2\theta}\left[2d_j\cos^j\theta f\partial_r+\partial_r\left(fd_j\right)\cos^j\theta\right]\nonumber\\
&+\frac{imf}{r^4\sin^3\theta}e_j\left(j\cos^{j-1}\theta\sin^2\theta+\cos^{j+1}\theta\right)-\frac{2imf}{r^4\sin^2\theta}e_j\cos^j\theta\partial_\theta\nonumber\\
&+\frac{f}{r^2}\left[\partial_{r}\left(fc_j\right)\cos^j\theta\partial_\theta+2fc_j\cos^j\theta\partial_{r\theta}^2\right]+\frac{f^2}{r^2\sin\theta}c_j\partial_\theta\left(\cos^j\theta\sin\theta\right)\partial_r\,.\label{kg1first}
\end{align}
\end{widetext}
Note that the summations over $j$ are implicitly assumed in each term in Eq.~\eqref{kg1first}.

To proceed, we adopt the ansatz
\begin{equation}
\psi_{m,\omega}=\sum_{l'=|m|}^\infty P_{l'}^m(x)R_{l',m}(r)\label{expansionwavefn}
\end{equation}
for the wave function and operate it using the operator \eqref{D01}. Note that we have defined $x=\cos\theta$ for simplicity. The associated Legendre functions $P_{l}^m(x)$ are the angular basis of the zeroth order operator \eqref{KGoperator1}. More explicitly, at the zeroth order, the master equation for wave functions satisfying Eq.~\eqref{expansionwavefn} can be factorized into a series of radial differential equations, each containing a particular mode of multiple number $l$ in that ansatz. Once the deformations are included, the master equation becomes non-separable as it can no longer be factorized into equations of particular modes. However, since the off-diagonal terms, i.e., modes with multiple numbers $l$ different from that of the zeroth order terms, say, $l_0$, are only of order $\epsilon$, one can project out the off-diagonal terms and focus only on the corrections on the zeroth order equation \cite{Cano:2020cao}.

The procedure that we have just mentioned above can be illustrated explicitly as follows. We first rewrite the ansatz \eqref{expansionwavefn} as
\begin{equation}
\psi_{m,\omega}=P_{l_0}^m(x)R_{l_0,m}(r)+\epsilon\sum_{l\ne l_0}P_{l}^m(x)R_{l,m}(r)\,,\label{revision1}
\end{equation}
where it is made explicitly that the off-diagonal terms ($l\ne l_0$) are of order $\epsilon$. Inserting Eq.~\eqref{revision1} into the operator \eqref{D01} and keeping terms up to $O(\epsilon)$, we get 
\begin{align}
&\mathcal{D}_{m,\omega}^2\psi_{m,\omega}\nonumber\\
=&\,\left(\mathcal{D}_{(0)m,\omega}^2+\epsilon\mathcal{D}_{(1)m,\omega}^2\right)\left[P_{l_0}^m(x)R_{l_0,m}(r)\right]\nonumber\\
&+\epsilon\sum_{l\ne l_0}\mathcal{D}_{(0)m,\omega}^2\left[P_{l}^m(x)R_{l,m}(r)\right]\,.\label{revision2}
\end{align}
Note that the associated Legendre functions $P_{l}^m(x)$ are the angular basis of the operator $\mathcal{D}_{(0)m,\omega}^2$, i.e., they are eigenfunctions of the last term of the operator \eqref{KGoperator1}. The projection method is performed by taking an inner product with a $P_{l_0}^m(x)$:
\begin{align}
&\int_{-1}^1dxP_{l_0}^m(x)\mathcal{D}_{m,\omega}^2\psi_{m,\omega}\nonumber\\
=&\,\int_{-1}^1dxP_{l_0}^m(x)\left(\mathcal{D}_{(0)m,\omega}^2+\epsilon\mathcal{D}_{(1)m,\omega}^2\right)\left[P_{l_0}^m(x)R_{l_0,m}(r)\right]\,.\label{revision3}
\end{align}
Note that the off-diagonal terms in Eq.~\eqref{revision2} have been projected out because $P_l^m(x)$ with the same $m$ are orthogonal. For simplicity, we replace the notation $l_0$ in Eq.~\eqref{revision3} with $l$ from now on.

By the normalization condition
\begin{align}
\int_{-1}^1dxP^m_l(x)P^m_k(x)&=\mathcal{N}_{lm}\delta_{lk}\,,\nonumber\\\textrm{where}&\quad\mathcal{N}_{lm}\equiv\frac{2(l+m)!}{(2l+1)(l-m)!}\,,
\end{align}
the zeroth order operator in Eq.~\eqref{revision3} can be written as
\begin{align}
&\frac{1}{\mathcal{N}_{lm}}\int_{-1}^{1}dxP^m_l(x)\mathcal{D}^2_{(0)m,\omega}\left[P_{l}^m(x)R_{l,m}(r)\right]
\nonumber\\=&\,\left[-\omega^2+\frac{f}{r^2}l(l+1)-\frac{f}{r^2}\partial_r\left(r^2f\partial_r\right)\right]R_{l,m}\,.
\end{align}
By inserting $f(r)=1-2M/r$, one gets the Regge-Wheeler equation with spin $s=0$, namely, the master equation for massless scalar fields propagating in the Schwarzschild spacetime.

For the first order operator, we find it convenient to define the following coefficients:
\begin{align}
a^j_{lm}&=\frac{m^2}{\mathcal{N}_{lm}}\int_{-1}^1\frac{x^j\left({P_l^m}\right)^2}{1-x^2}dx\,,\\
b^j_{lm}&=\frac{1}{\mathcal{N}_{lm}}\int_{-1}^1x^j\left({P_l^m}\right)^2dx\,,\\
c^j_{lm}&=\frac{1}{\mathcal{N}_{lm}}\int_{-1}^1x^j{P_l^m}\left[\left(1-x^2\right)\partial_x^2-2x\partial_x\right]{P_l^m}dx\,,\end{align}
\begin{align}
d^j_{lm}&=\frac{1}{\mathcal{N}_{lm}}\int_{-1}^1{P_l^m}\left(1-x^2\right)\left(\partial_xx^j\right)\left(\partial_x{P_l^m}\right)dx\,,\\
e^j_{lm}&=\frac{-1}{\mathcal{N}_{lm}}\int_{-1}^1dxx^j{P_l^m}\sqrt{1-x^2}\partial_x{P_l^m}\,,\\
f^j_{lm}&=\frac{1}{\mathcal{N}_{lm}}\int_{-1}^1dx\left({P_l^m}\right)^2\left[\frac{x^{j+1}}{\sqrt{1-x^2}}-\sqrt{1-x^2}\partial_xx^j\right]\,,
\end{align}
\begin{align}
g^j_{lm}&=\frac{m}{\mathcal{N}_{lm}}\int_{-1}^1\frac{x^j\left({P_l^m}\right)^2dx}{1-x^2}\,,\\
h^j_{lm}&=\frac{m}{\mathcal{N}_{lm}}\int_{-1}^1\frac{\left({P_l^m}\right)^2dx}{\left(1-x^2\right)^{3/2}}\left[jx^{j-1}\left(1-x^2\right)+x^{j+1}\right]\nonumber\\
&+\frac{2m}{\mathcal{N}_{lm}}\int_{-1}^1\frac{x^j{P_l^m}\left(\partial_x{P_l^m}\right)dx}{\sqrt{1-x^2}}\,.
\end{align}
By using integrations by part, it can be shown that these coefficients satisfy the following relations
\begin{equation}
2e^j_{lm}+f^j_{lm}=0\,,\qquad h^j_{lm}=0\,.\label{efh}
\end{equation}
Also, we have
\begin{align}
c^j_{lm}&=-l(l+1)b^j_{lm}+a^j_{lm}\,,\nonumber\\
d^0_{lm}&=0\,,\nonumber\\
a^j_{lm}&=b^j_{lm}=c^j_{lm}=d^j_{lm}=g^j_{lm}=0\quad\textrm{if }j\textrm{ is odd,}
\end{align}
and
\begin{equation}
e^j_{lm}=f^j_{lm}\qquad\textrm{if }j\textrm{ is even.}\label{efevenidentity}
\end{equation}
The identity \eqref{efevenidentity} and the first identity of Eq.~\eqref{efh} directly imply $e^j_{lm}=0$ if $j$ is even. 

With the above coefficients, the first order operator can be expressed as
\begin{widetext}
\begin{align}
&\frac{1}{\mathcal{N}_{lm}}\int_{-1}^1dx{P_l^m}\mathcal{D}_{(1)m,\omega}^2\left[P_{l}^m(x)R_{l,m}(r)\right]\nonumber\\=&\,\frac{f}{r^2}\left[a^j_{lm}\left(A_j-D_j\right)-c^j_{lm}\left(A_j-C_j\right)-\frac{d^j_{lm}}{2}\left(A_j+B_j-C_j+D_j\right)+e^j_{lm}\partial_r\left(fc_j\right)-ig^j_{lm}\partial_r\left(fd_j\right)\right]R_{l,m}\nonumber\\&-\frac{f}{r^2}b^j_{lm}\left(A_j-B_j\right)\partial_r\left(r^2f\partial_r\right)R_{l,m}-\frac{f^2}{2r^2}\left[b^j_{lm}r^2\left(A_j'-B_j'+C_j'+D_j'\right)+4ig^j_{lm}d_j+\frac{4i\omega r^2b^j_{lm}a_j}{f}\right]\partial_rR_{l,m}\nonumber\\
&-\frac{i\omega fb^j_{lm}}{r}\left(2a_j+ra_j'\right)R_{l,m}\,.\label{firstorderproject}
\end{align}
\end{widetext}
Again, the summations over $j$ are implicitly assumed, and the prime denotes the derivatives with respect to $r$.

\subsection{Effective potential}
As we have mentioned, at the zeroth order, the master equation can be written as a Schr\"odinger-like form, which contains an effective potential. It can also be done in the presence of spacetime deformations if only taking $\epsilon$-order effects into account. After recasting the master equation into the Schr\"odinger-like form and determining the effective potential in the presence of deformations, we can directly compare how the deformations modify the effective potential and, in turn, the QNMs themselves.

In order to rewrite the master equation into a Schr\"odinger-like form, we define the tortoise radius $r_*$ that satisfies
\begin{equation}
\frac{dr}{dr_*}=f(r)\left[1+\frac{\epsilon}{2}b^j_{lm}\left(A_j-B_j\right)\right]\,.
\end{equation}
Then, we redefine a new radial wave function $\Psi_{l,m}$ relating to the original radial wave function $R_{l,m}$ as
\begin{equation}
R_{l,m}=\frac{\Psi_{l,m}}{r}\left[1+\frac{\epsilon}{4}b^j_{lm}\left(A_j-B_j\right)-\epsilon\int dr\frac{Z_{lm}(r)}{4r^2}\right]\,,
\end{equation}
where
\begin{align}
Z_{lm}(r)&\equiv b^j_{lm}r^2\left(A_j'-B_j'+C_j'+D_j'\right)\nonumber\\&+ 4i g^j_{lm}d_j+\frac{4i\omega r^2b^j_{lm}a_j}{f}\,.
\end{align}
The master equation can then be rewritten in the Schr\"odinger-like form
\begin{equation}
\partial_{r_*}^2\Psi_{l,m}+\omega^2\Psi_{l,m}=V_{\textrm{eff}}(r)\Psi_{l,m}\,,
\end{equation}
where the modified effective potential reads
\begin{widetext}
\begin{align}
V_{\textrm{eff}}(r)=&\,l(l+1)\frac{f}{r^2}+\frac{f}{r}\frac{df}{dr}\left[1+\epsilon b^j_{lm}\left(A_j-B_j\right)\right]+\epsilon\Bigg\{\frac{f}{r^2}\Big[a^j_{lm}\left(A_j-D_j\right)-c^j_{lm}\left(A_j-C_j\right)-\frac{d^j_{lm}}{2}\left(A_j+B_j-C_j+D_j\right)\nonumber\\&+\,e^j_{lm}\partial_r\left(fc_j\right)\Big]+\frac{1}{4r^2}\frac{d}{dr_*}\left[b^j_{lm}r^2\frac{d}{dr_*}\left(A_j-B_j+C_j+D_j\right)\right]-\frac{b^j_{lm}}{4}\frac{d^2}{dr_*^2}\left(A_j-B_j\right)\Bigg\}  \,.\label{eq:Veff}
\end{align}
\end{widetext}
Note that only the coefficients $a^j_{lm}$, $b^j_{lm}$, $c^j_{lm}$, $d^j_{lm}$, and $e^j_{lm}$ appear in the effective potential $V_{\textrm{eff}}(r)$. In addition, although the right-hand side of Eq.~\eqref{firstorderproject} depends on the frequency $\omega$, the effective potential \eqref{eq:Veff} does not. It is then clear from Eq.~\eqref{eq:Veff} how the deformation functions in the metric modify the effective potential of the master equation. Finally, we would like to mention that $g_{r\theta}$, i.e., the metric function $c_j(r)$, is the only off-diagonal metric component that enters the effective potential. However, it does not mean that other off-diagonal metric components are not physical. In fact, as we have shown in sec.~\ref{sec.spacetime}, the deformed spacetime is non-circular due to the presence of the metric functions in other off-diagonal components. Therefore, these off-diagonal metric components cannot be removed by simply using coordinate transformations. It just happens that they do not contribute to the effective potential due to our neglect of higher-order deformation effects. Note that, although the metric functions that encode the spacetime deformations can be quite arbitrary, in the rest of this paper, we will assume the deformations to be mild enough such that the effective potential \eqref{eq:Veff} still has a single peak for simplicity.

\section{Eikonal limit}\label{sec.eikonal}
As we have mentioned, the correspondence between eikonal QNMs and photon geodesics, in particular, the photon sphere around black holes, is tightly related to the fact that in the eikonal limit, the peak of the effective potential is located precisely at the photon sphere. In the previous section, we have obtained the effective potential for the radial master equation of QNMs. It allows us to examine how the peak of the effective potential would be altered in the presence of spacetime deformations.

In the case of static and spherically symmetric spacetimes, the effective potential contains only the multiple number $l$, not the azimuthal number $m$. Therefore, the eikonal limit corresponds to the limit where $l$ goes to infinity ($l\gg1$). However, in the presence of spacetime deformations, the effective potential depends on $m$, and thus, the behaviors of high-frequency modes could be different for different values of $m$. In this section, we will first focus on two special cases: $|m|=l$ and $m=0$, then discuss how the effective potential would behave in the eikonal limit for these cases separately. The general discussions on arbitrary $m$ will be presented at the end of this section.

\subsection{$|m|=l$}
We first consider the case with $|m|=l$. Note that the associated Legendre functions with $m=l$ can be expressed as
\begin{align}
P_0^0&=1\,,\nonumber\\P_l^l(x)&=\left(-1\right)^l\frac{\left(2l-1\right)!}{2^{l-1}\left(l-1\right)!}\left(1-x^2\right)^{l/2}\,,\quad\textrm{for $l\ge1$}\,.
\end{align}
On the other hand, for the case with $m=-l$, we have
\begin{equation}
P_l^{-l}(x)=\frac{\left(2l-1\right)!}{2^{l-1}\left(l-1\right)!\left(2l\right)!}\left(1-x^2\right)^{l/2}\,.
\end{equation}
Using the above expressions, one can directly compute the coefficients $a^j_{ll}$, $a^j_{l-l}$, $b^j_{ll}$, $b^j_{l-l}$, $c^j_{ll}$, $c^j_{l-l}$, $d^j_{ll}$, and $d^j_{l-l}$. When $j$ is even, i.e., when $j$ equals $2k$ with $k$ being non-negative integers, these coefficients are
\begin{align}
a^{2k}_{ll}&=a^{2k}_{l-l}=\frac{l\left(2l+1\right)C^{l+k}_k}{2C^{2l+2k}_{2k}}\,,\\
b^{2k}_{ll}&=b^{2k}_{l-l}=\frac{\left(2l+1\right)C^{l+k}_k}{\left(2l+2k+1\right)C^{2l+2k}_{2k}}\,,\\
c^{2k}_{ll}&=c^{2k}_{l-l}=\frac{l\left(2l+1\right)\left(2k-1\right)C^{l+k}_k}{2\left(2l+2k+1\right)C^{2l+2k}_{2k}}\,,\\
d^{2k}_{ll}&=d^{2k}_{l-l}=-\frac{2kl\left(2l+1\right)C^{l+k}_k}{\left(2l+2k+1\right)C^{2l+2k}_{2k}}\,,
\end{align}
where $C^{a}_b\equiv a!/[b!(a-b)!]$ is the combination number. When $j$ is odd, these coefficients are identically zero.

As for the coefficients $e^j_{ll}$ and $e^j_{l-l}$, one can use Stirling's approximation to get their asymptotic expressions in the eikonal limit ($l\gg1$):
\begin{equation}
e^{j}_{ll}\approx e^{j}_{l-l}\approx\frac{2(2k+1)(2k)!}{4^{k+1}k!}l^{-k}\,,
\end{equation}
when $j$ is odd ($j=2k+1$), and $e^{j}_{ll}=e^{j}_{l-l}=0$ when $j$ is even.

According to the above expressions, in the eikonal limit ($l\gg1$) for the case with $|m|=l$, the dominant coefficients are $a^0_{ll}$ and $a^0_{l-l}$, approximated as
\begin{align}
a^0_{ll}=a^0_{l-l}\approx l^2\,.
\end{align}
Therefore, in this case, the effective potential \eqref{eq:Veff} can be approximated as 
\begin{equation}
V_{\textrm{eff}}(r)\approx l^2\frac{f}{r^2}\left[1+\epsilon\left(A_0-D_0\right)\right]\,.\label{potentialll}
\end{equation}
Notice that only the deformation functions with $j=0$ in the $g_{tt}$ and $g_{\varphi\varphi}$ components dominate the effective potential of eikonal modes with $|m|=l$.

\subsection{$m=0$}
For the case with $m=0$, the associated Legendre function reduces to the Legendre function $P_l$. In the large $l$ limit, we have
\begin{equation}
P_l\left(\cos\theta\right)=\frac{2}{\sqrt{2\pi l\sin\theta}}\cos\left[\left(l+\frac{1}{2}\right)\theta-\frac{\pi}{4}\right]+\mathcal{O}\left(l^{-3/2}\right)\,.
\end{equation}
Plugging the formula above into coefficients, when $j$ is even ($j=2k$), we have
\begin{align}
b^j_{l0}&\approx\frac{1}{4^k}C^{2k}_k\,,\label{bl0}\\
c^j_{l0}&\approx-\frac{1}{4^k}C^{2k}_kl^2\,,\label{cl0}\\
d^j_{l0}&\approx\frac{k}{4^k}C^{2k}_k\,,
\end{align}
while when $j$ is odd ($j=2k+1$), these coefficients are identically zero. Clearly, in the eikonal limit, the coefficients $c^j_{l0}$ with even $j$ dominate over the coefficients $b^j_{l0}$ and $d^j_{l0}$ because $c^j_{l0}$ scale quadratically while $b^j_{l0}$ and $d^j_{l0}$ scale linearly in $l$.

The estimation of the coefficients $a^j_{l0}$ can be achieved by identity $c^j_{lm}=-l(l+1)b^j_{lm}+a^j_{lm}$. It can be directly seen from Eqs.~\eqref{bl0} and \eqref{cl0} that the coefficients $a^j_{l0}$ are subdominant compared with $c^j_{l0}$. 

Finally, the estimation of the coefficients $e^j_{l0}$ seems more involved. At this point, we will only exhibit that the coefficients $e^j_{l0}$ are also subdominant compared with $c^j_{l0}$. We shall only focus on the cases in which $j$ is odd ($j=2k+1$) because $e^j_{l0}=0$ for even $j$. We can write
\begin{align}
|e^j_{l0}|&=\frac{2}{\mathcal{N}_{l0}}\left|\int_{0}^1dxx^{2k+1}\sqrt{1-x^2}P_l(x)\partial_xP_l(x)\right|\nonumber\\
&\le\frac{2}{\mathcal{N}_{l0}}\int_{0}^1dxx^{2k+1}\sqrt{1-x^2}\left|P_l(x)\right|\left|\partial_xP_l(x)\right|\nonumber\\
&\le\frac{4}{\mathcal{N}_{l0}\pi}\sqrt{\frac{l+\frac{2}{3}}{l+\frac{1}{2}}}\int_0^1dx\frac{x^{2k+1}}{\left(1-x^2\right)^{1/4}}\nonumber\\
&=\frac{k!\Gamma\left(3/4\right)}{\pi\Gamma\left(k+7/4\right)}\sqrt{\frac{l+\frac{2}{3}}{l+\frac{1}{2}}}\left(2l+1\right)\,.\label{estimatione}
\end{align}
During the above estimation, we have used the improved version of Bernstein's inequality \cite{Antonov1981,Lorch1993}
\begin{align}
\left|P_l(x)\right|<\sqrt{\frac{2}{\pi\left(l+1/2\right)}}&\frac{1}{\left(1-x^2\right)^{1/4}}\,,\nonumber\\&\textrm{for }l\ge0\textrm{ and }x\in[-1,1]\,.
\end{align}
To estimate the derivative of Legendre polynomials, we have used the following inequality \cite{Antonov:2010}
\begin{align}
\left|\left(1-x^2\right)\partial_xP_l(x)\right|&<\sqrt{\frac{2}{\pi}\left(l+2/3\right)}\,,\nonumber\\&\textrm{for }l\ge0\textrm{ and }x\in[-1,1]\,.
\end{align}

In the eikonal limit, the right-hand side of the inequality \eqref{estimatione}, which acts as an upper bound of $|e^j_{l0}|$, scales as $O(l)$. Therefore, in this limit, the coefficients $e^j_{l0}$ are subdominant compared with the coefficients $c^j_{l0}$ that scale as $O(l^2)$. As a consequence, in the eikonal limit, the effective potential for QNMs with $m=0$ can be approximated as
\begin{equation}
V_{\textrm{eff}}(r)\approx l^2\frac{f}{r^2}\left[1+\epsilon\sum_{k=0}^\infty\frac{1}{4^k}C^{2k}_k\left(A_{2k}-C_{2k}\right)\right]\,.\label{potentialm0}
\end{equation}

\subsection{Generic $m$}

For generic cases with arbitrary $m$, we may utilize the classical limit of the 3-$J$ symbol, i.e., the integration of the products between three arbitrary spherical harmonics $Y_{lm}$, $Y_{s m_s}$, and $Y_{(l+\delta l) (m+m_s)}^*$ over a $2$-sphere, to generate the coefficients in the potential \cite{2010IJQC..110..731R},
\begin{align}
&\lim_{l \to \infty}\int Y_{lm} Y_{s m_s} Y_{(l+\delta l) (m+m_s)}^* d\Omega  \nonumber\\
\approx &\sqrt{\frac{2s+1}{4 \pi}} d^s_{m_s,\delta l} \left(\frac{\pi }{2}\right) d^s_{m_s,\delta l} \left(\cos^{-1}\alpha\right) \,,
\end{align}
where $Y_{lm} (\theta,\varphi) \equiv \sqrt{\frac{(2l+1) (l-m)!}{4\pi (l+m)!}} P_l^m (\cos\theta) e^{im\varphi}$ is the spherical harmonics, $\Omega$ is the measure over a $2$-sphere, $d^s_{m_s,\delta l} (\beta) \equiv \int Y_{s m_s}^* \mathcal{R}(\beta)Y_{s \delta l} \,d\Omega$ is the Wigner d matrix with a pitch $\mathcal{R}$ of angle $\beta$, and $\alpha \equiv (2m+m_s)/(2l+\delta l +1)$ is the cosine of the rotation angle that will be related to the inclination angle of the corresponding orbit.

By relating $x^j$ to $Y_{s m_s}$ and fixing $\delta l = m_s = 0$, the coefficient $b^j_{lm}$ turns out to be $1, 0, (1-\alpha^2)/2, 0, 3(1-\alpha^2)^2/8...$, for $j= 0,1,2...$, with the generating function $G(z) = (1-(1-\alpha^2)z^2)^{-1/2}$. Thus, we obtain the generic formulas of the coefficients,
\begin{align}
a_{(2k)lm} &\approx
l^2 \alpha ^2 \left( 1-\alpha^2 \right)^k 4^{-k} C^{2k}_k  \nonumber\\
&\times \, _2F_1 \left( 1,k+\frac{1}{2};k+1;1-\alpha^2 \right)  \,,  \nonumber\\
b_{(2k)lm} &\approx
\left(1-\alpha^2\right)^k 4^{-k} C^{2k}_k \,,  \nonumber\\
c_{(2k)\,lm} &\approx
a_{(2k)\,lm} - l^2 b_{(2k)\,lm}  \,,  \nonumber\\
d_{(2k)lm} &\approx
\left(1-\alpha ^2\right)^{k-1} 4^{-k} C^{2k}_k \left( k - k\left( 2k+1 \right) \alpha^2 \right)  \,,  \nonumber\\
e_{(2k+1)lm} &\approx
\left(1-\alpha^2\right)^k 4^{-k-1} C^{2k+2}_{k+1}  \nonumber\\
&\times \left( (k+1) \, _2F_1\left(-\frac{1}{2},k+\frac{1}{2};k+1;1-\alpha ^2\right) \right.  \nonumber\\
&\left.
- \frac{1-\alpha^2}{2} \, _2F_1\left(\frac{1}{2},k+\frac{3}{2};k+2;1-\alpha ^2\right) \right)  \,,  \label{eq:abcde_general}
\end{align}
where $F$ is the hypergeometric function. Coefficients of opposite parity (odd for $a\sim d$ and even for $e$) vanish exactly.

The formulas above match the leading order coefficients obtained in previous subsections but not the higher-order ones such as $c_{2ll}$ or $e_{jl0}$ since the approximation deployed in this subsection contains $o(1)$ subtraction when converting $Y_{s0}$ to $x^j$.
Nonetheless, we are certain that only $a_{jlm}$ and $c_{jlm}$ survive the eikonal limit regardless of $m$ since they are the only terms of order $l^2$ in $m=0 \lor \pm l$ cases discussed previously, or when $\alpha$ is away from $0$ ($m=0$) and $1$ ($|m| = l$).

For completeness, let us write down the effective potential for eikonal QNMs of arbitrary $m$
\begin{widetext}
\begin{align}
V_{\textrm{eff}}(r) &\approx l^2 \frac{f}{r^2} \left\{ 1 + \epsilon \sum_{k=0}^\infty \left( 1-\alpha^2 \right)^{k} 4^{-k} C^{2k}_k \left[ \alpha^2 \, _2F_1 \left(1,k+\frac{1}{2};k+1;1-\alpha^2 \right) \left( A_{2k} - D_{2k} \right)  \right.\right.  \nonumber\\
&+ \left.\left. \frac{1-\alpha^2}{2k+2} \, _2F_1 \left(1,k+\frac{1}{2};k+2;1-\alpha^2 \right)  \left( A_{2k} - C_{2k} \right) \right] \right\} \,,\label{eq:potential_m_generic}
\end{align}
where we have applied the identity $\alpha^2 \, _2F_1 \left(1,k+\frac{1}{2};k+1;1-\alpha^2 \right) + \frac{1-\alpha^2}{2k+2} \, _2F_1 \left(1,k+\frac{1}{2};k+2;1-\alpha^2 \right) = 1$.
\end{widetext}

\section{Eikonal correspondence with photon orbits}\label{sec.geodecorr}
Roughly speaking, the correspondence between eikonal QNMs of spherically symmetric black holes and the photon sphere originates from the location of the peak of the effective potential in the master equation for eikonal QNMs, which is at the photon sphere around the black hole. In this section, we will extend the investigation to the deformed Schwarzschild black hole \eqref{parametric}. We will first consider circular photon orbits, in which photons undergo planar motion parallel to the equatorial plane $\theta=\pi/2$. It turns out that these orbits still exist in this spacetime, and there is a correspondence between these circular orbits and the eikonal QNMs with $|m|=l$. Then, we will consider polar orbits that are \textit{nearly circular}, i.e., with slightly varying radii. However, these orbits are periodic and cross the poles ($|x|=1$) repeatedly. We will show that the same as the cases for generic orbits, the peak of the effective potential coincides with the \textit{averaged} radius of these nearly circular orbits along one complete period. 

\subsection{Circular photon orbits}   

We first reconsider the deformed Schwarzschild metric \eqref{parametric} and rewrite it in the following form:
\begin{align}
ds^2=&\,g_{tt}dt^2+g_{rr}dr^2+g_{\theta\theta}d\theta^2+g_{\varphi\varphi}d\varphi^2\nonumber\\
&+2g_{tr}drdt+2g_{t\theta}d\theta dt+2g_{r\varphi}drd\varphi+2g_{\theta\varphi}d\theta d\varphi\nonumber\\&+2g_{r\theta}drd\theta\,.\label{generalmetricexpression}
\end{align}
According to Eq.~\eqref{parametric}, the diagonal components of the metric have non-vanishing zeroth order parts, i.e., the Schwarzschild metric, plus the first order terms in $\epsilon$. On the other hand, the off-diagonal parts are $O(\epsilon)$ by themselves.

Then, we consider the geodesic equations of massless particles moving in the spacetime \eqref{generalmetricexpression}. The axisymmetry and stationarity lead to two constants of motion,
\begin{align}
E&=-g_{tt}\dot{t}-g_{tr}\dot{r}-g_{t\theta}\dot\theta\,,\\
L_z&=g_{\varphi\varphi}\dot\varphi+g_{r\varphi}\dot{r}+g_{\theta\varphi}\dot\theta\,,
\end{align}
where $E$ and $L_z$ are the energy and the azimuthal angular momentum of particles. The dot denotes the derivative with respect to the affine parameter $\lambda$. The above equations can be solved to get
\begin{align}
\dot{t}&=-\frac{E+g_{tr}\dot{r}+g_{t\theta}\dot\theta}{g_{tt}}  \label{eq:tdot}  \,,\\
\dot{\varphi}&=\frac{L_z-g_{r\varphi}\dot{r}-g_{\theta\varphi}\dot\theta}{g_{\varphi\varphi}}\,. \label{eq:phidot}
\end{align}
Expanding up to the first order in $\epsilon$, the constraint equation of photon geodesics can be written as
\begin{equation}
\frac{E^2}{g_{tt}}+\frac{L_z^2}{g_{\varphi\varphi}}+g_{rr}\dot{r}^2+g_{\theta\theta}\dot\theta^2+2g_{r\theta}\dot{r}\dot\theta=0\,.  \label{eq:constraint}
\end{equation}
The circular orbits are defined by photons moving at a constant radius $r$ and on a fixed plane parallel to the equatorial plane{\footnote{In general, these photon orbits are not on the equatorial plane because the spacetime is equator reflection asymmetric.}}. These photon orbits are then associated with $\dot{r}=\ddot{r}=\dot{\theta}=\ddot{\theta}=0$ for all affine time $\lambda$, thus satisfying the following constraint equation
\begin{equation}
\frac{1}{g_{tt}}+\frac{b^2}{g_{\varphi\varphi}}=0\,,\label{v0}
\end{equation}
where $b\equiv L_z/E$ is the impact parameter of the photon orbits.

Using the geodesic equation
\begin{equation}
\frac{d}{d\lambda}\left(g_{\mu\nu}\dot{x}^\nu\right)=\frac{1}{2}\left(\partial_{\mu}g_{\alpha\beta}\right)\dot{x}^\alpha\dot{x}^\beta\,,  \label{eq:geodesic}
\end{equation}
requiring $\dot{r}=\dot{\theta}=\ddot{r}=\ddot{\theta}=0$, and taking its radial and polar angle components, we get
\begin{align}
\frac{1}{g_{tt}^2}\partial_rg_{tt}+\frac{b^2}{g_{\varphi\varphi}^2}\partial_rg_{\varphi\varphi}&=0\,,\label{vr}\\
\frac{1}{g_{tt}^2}\partial_\theta g_{tt}+\frac{b^2}{g_{\varphi\varphi}^2}\partial_\theta g_{\varphi\varphi}&=0\,.\label{vtheta}
\end{align}
In principle, one can obtain the associated radius $r$, the polar angle $\theta$, and the impact parameter $b$ of the orbits by simultaneously solving Eqs.~\eqref{v0}, \eqref{vr}, and \eqref{vtheta}. Now, we will show that the radius $r$ of these orbits must be at the peak of Eq.~\eqref{potentialll}, i.e., the potential corresponding to the eikonal QNMs with $|m|=l$. It can be done without having explicit forms of the deformation functions $A_j(r)$ and $D_j(r)$.

We first start with Eq.~\eqref{vtheta}. In the presence of spacetime deformations, Eq.~\eqref{vtheta} implies that circular orbits would lie on planes with $\theta=\pi/2+O\left(\epsilon\right)$.\footnote{Only deformations of odd parity along $\theta$ contribute.} By eliminating $b^2$ in Eqs.~\eqref{v0} and \eqref{vr}, one finds that the radius $r$ of circular photon orbits can be determined by
\begin{equation}
\partial_r\left(\frac{g_{tt}}{g_{\varphi\varphi}}\right)=0\,,
\end{equation}
which, up to the first order in $\epsilon$, can be written as
\begin{equation}
\partial_r\left[\frac{f\left(1+\epsilon A_0\right)}{r^2\left(1+\epsilon D_0\right)}\right]\approx\partial_r\left[\frac{f}{r^2}\left(1+\epsilon A_0-\epsilon D_0\right)\right]=0\,.\label{radiuscircularml}
\end{equation}
Note that no $\theta$ dependence appears in the equation because $\epsilon\cos\theta\sim O\left(\epsilon^2\right)$ and $\sin\theta\approx1+O\left(\epsilon^2\right)$, as the only terms depending on $\theta$, receive no correction. According to Eq.~\eqref{radiuscircularml}, one sees that the radius of circular photon orbits follows the same equation that determines the peak of the effective potential \eqref{potentialll}. Therefore, the correspondence between eikonal QNMs with $|m|=l$ and circular photon orbits holds for deformed Schwarzschild black holes.

\subsection{Polar orbits}
After constructing the correspondence between eikonal QNMs and circular photon orbits for deformed Schwarzschild black holes, we then switch gear and consider the polar photon orbits and see whether an analog correspondence for these orbits also exists. The discussion about photon orbits with arbitrary inclinations will be exhibited later. 

It is well known that the spherical photon orbits around Schwarzschild black holes are always perfect circles. More explicitly, each of them is a light ring with a radius $r=3M$, the union of which forms a photon sphere. After the spacetime is deformed, most of the circular orbits would become neither circular nor planar, except for the special orbits that remain parallel to the equatorial plane, i.e., the orbits discussed in the previous subsection. For those orbits that receive non-circular deformations, since the spacetime deformations are $O(\epsilon)$, the deviations of the orbits from circular and planar motions also remain $O(\epsilon)$. For these generic ``light rings'' with non-zero inclination angles to the axis of spacetime symmetry, the light ring radius should pick up a non-constant deformation. Therefore, the correspondence between the potential peak of the QNMs, which is a constant for a particular mode, and the light ring radius, seems ill defined. Nevertheless, let us proceed and solve the orbit, in the hope of finding a correspondence that permits physical interpretations.

Let us consider the geodesic equations for polar orbits first. Natural conditions would be $\dot r \sim O(\epsilon)$ as we focus on orbits around the photon sphere, and $L_z \sim O(\epsilon)$. Therefore, Eqs.~\eqref{eq:tdot}, \eqref{eq:phidot}, and the constraint equation
become
\begin{align}
\dot t &= \frac{E+g_{t\theta}\dot\theta
}{-g_{tt}}  \,,\\
\dot \varphi &= \frac{L_z-g_{\theta\varphi}\dot \theta
}{g_{\varphi\varphi}}  \,,\\
0 &=  \frac{E^2}{-g_{tt}} - \frac{\mathfrak{A}^2 \dot{\theta}^2 + L_z^2}{g_{\varphi\varphi}} + O(\epsilon^2)  \,,
\end{align}
where $\mathfrak{A} \equiv \sqrt{g_{\theta\theta} g_{\varphi\varphi} - g_{\theta\varphi}^2}$ is the area element of $\theta$, $\varphi$, and is everywhere semi-positive-definite.
For the polar orbit to cross the poles ($g_{\varphi\varphi} = 0$) at a finite $\dot\theta$, in addition to the vanishing of $L_z$, the following condition $\mathfrak{A}^2 = O(g_{\varphi\varphi})$ around the poles is also mandatory \footnote{The condition ensures the topology of $S^2$.}, indicating that $g_{\varphi\varphi}^{-1}\mathfrak{A}^2$ receives $O(\epsilon^2)$ corrections everywhere. Unfortunately, $\dot\varphi$ could be large, and extra care must be taken.
We now turn our attention to the $r$ component of Eq.~\eqref{eq:geodesic}:
\begin{align}
\frac{d}{d\lambda}\left( g_{rr}\dot r \right)
&= \frac{E^2}{2g_{tt}} \partial_r\ln \left|\frac{g_{tt}}{g_{\theta\theta}}\right| - \frac{d}{d\lambda}\left( \left(g_{r\theta} - g_{r\varphi} \frac{g_{\theta\varphi}}{g_{\varphi\varphi}}\right) \dot\theta \right)  \nonumber\\
&+ \left( g_{t\theta} \partial_r \ln \left|\frac{g_{t\theta}}{g_{tt}}\right| + g_{tr} \partial_\theta \ln \left|\frac{g_{tt}}{g_{tr}}\right| \right) \dot t \dot\theta + O(\epsilon^2)  \,.\label{rrdpolar}
\end{align}

As we have just mentioned, these deformed ``light rings" do not have a constant radius $r$. However, these orbits have to be periodic as they pass through the poles and can be regarded as a class of limit cycles in the phase space. It allows us to integrate Eq.~\eqref{rrdpolar} along a closed loop for one period in $\lambda$ direction, i.e., $\oint d\lambda=\oint \dot\theta^{-1} d\theta$. For detailed reasons shown later along with the discussion for generic orbital inclinations, only the first term on the right-hand side of Eq.~\eqref{rrdpolar} dominates the integration. Therefore, we get
\begin{align}
o(\epsilon)&\propto\int_{0}^{2\pi}d\theta\partial_r\ln{\left|\frac{g_{tt}}{g_{\theta\theta}}\right|}\nonumber\\
&=\int_0^{2\pi}d\theta\partial_r\left[\frac{f}{r^2}\left(1+\epsilon\left(A_j-C_j\right)\cos^j\theta\right)\right]\nonumber\\
&\propto\partial_r\left[\frac{f}{r^2}\left(1+\epsilon\sum_{k=0}^{\infty}\frac{1}{4^k}C^{2k}_k(A_{2k}-C_{2k})\right)\right]\,,\label{polarintegration}
\end{align}
where the dummy index $j$ stands for summation over all non-negative integers. Note that the integration runs from $\theta=0$ to $\theta=2\pi$, meaning that the trajectories go back and forth through the two poles. According to Eq.~\eqref{polarintegration}, one sees that the integrated equation, up to $O(\epsilon)$, is equal to the equation that determines the peak of the effective potential \eqref{potentialm0}. Later in the next subsection, we will show explicitly that the peak of the effective potential can be interpreted as an averaged radius of deformed light rings along a period.

\subsection{Orbits with generic inclinations}

Methods in the last subsection can also be applied to general cases with $L_z \neq 0$. The only difference is that the orbits no longer pass the poles, sparing us from dealing with coordinate singularities. Again, let us consider the $r$ component of Eq.~\eqref{eq:geodesic},
\begin{align}
\frac{d}{d\lambda}\left( g_{rr}\dot r \right)
&= \frac{E^2}{2g_{tt}} \left( \left( 1 + \mathfrak{K} \right) \partial_r\ln \left|\frac{g_{tt}}{g_{\varphi\varphi}}\right| - \mathfrak{K}\, \partial_r\ln \left|\frac{g_{tt}}{g_{\theta\theta}}\right| \right)  \nonumber\\
&+ \left( g_{t\theta} \partial_r \ln \left|\frac{g_{t\theta}}{g_{tt}}\right| + g_{tr} \partial_\theta \ln \left|\frac{g_{tt}}{g_{tr}}\right| \right) \dot t \dot\theta  \nonumber\\
&+ \left( g_{\theta\varphi} \partial_r \ln \frac{g_{\theta\varphi}}{g_{\varphi\varphi}} + g_{r\varphi} \partial_\theta \ln \frac{g_{\varphi\varphi}}{g_{r\varphi}} \right) \dot\theta \dot\varphi  \nonumber\\
&- \frac{d}{d\lambda}\left( g_{r\theta} \dot\theta \right)  \,,\label{eq:geodesic_r_even}
\end{align}
with $\mathfrak{K} \equiv E^{-2} g_{tt} g_{\theta\theta} \dot \theta^2 = -1 - L_z^2 E^{-2} g_{tt} g_{\varphi\varphi}^{-1}\,$. As we argued before, ``light rings'' have to be periodic and thus correspond to a class of limit cycles. Let us focus on the Lyapunov exponent of a limit cycle $r^*$ on a $\theta$ section
\begin{align}
\left\langle \frac{d}{d\lambda}\left( g_{rr}\dot r \right) \right\rangle
&= \left\langle \partial_r F(r^*,\theta) (r-r^*) \right\rangle  \nonumber\\
&= \partial_r F_0(r_P) \left\langle r-r^* \right\rangle + o(\epsilon)  \nonumber\\
&= \partial_r F_0(r_P) \left\langle r-r_P \right\rangle + \epsilon \left\langle F_1(r_P,\theta) \right\rangle + O(\epsilon^2)  \,,
\end{align}
where $\langle\quad\rangle \equiv \oint \dot\theta^{-1} d\theta$ is the integration over one revolution, with $r$, $r^*$, and $\dot\theta$ considered functions of $\theta$, $F(r,\theta)$ represents the right-hand side of Eq.~\eqref{eq:geodesic_r_even}, $F_n$ is the $n$-th order term of $F$ in $\epsilon$ with $F_0$ independent of $\theta$, and $r_P$ is the photon sphere radius of a Schwarzschild black hole. The first equality defines the Lyapunov exponent of the deformed limit cycle $\Lambda \equiv \sqrt{\partial_r F(r^*,\theta)}$, while the third is the $\epsilon$-expansion of $F$. The second equality comes from the largeness of the background Lyapunov exponent, suggesting $O(\epsilon)$ proximity of the deformed limit cycle to the Schwarzschild one. Therefore, we may utilize background $\theta$ to determine $r^*$, rendering the last three terms on the right-hand side of Eq.~\eqref{eq:geodesic_r_even} $o(\epsilon)$ after tracing over one revolution.\footnote{Two of these terms are the non-circularity introduced in section \ref{sec.spacetime}. The suppression of the non-circularity can be understood as the irrelevance of the wave function deformation at linear order.} The integrated equation can be interpreted as determining up to $O(\epsilon)$ the averaged photon ring radius if the orbit is closed radially up to $o(\epsilon)$.

The orbit up to $O(1)$ can be characterized by its inclination angle $\delta$ as
\begin{align}
\mathfrak{K} = - \left( 1-x^2 \right)^{-1} \left( \sin^2\delta -x^2 \right)  \,,\\
\dot x = \pm \sqrt{\frac{E^2 \left( \sin^2\delta-x^2 \right)}{-g_{tt} g_{\theta\theta}}}  \,,
\end{align}
with $x=\cos\theta$, and the integration of Eq.~\eqref{eq:geodesic_r_even} over one period is
\begin{align}
o(\epsilon) &= \int_{\mp\sin\delta}^{\pm\sin\delta}\left( \left( 1 + \mathfrak{K} \right) \partial_r\ln \left|\frac{g_{tt}}{g_{\varphi\varphi}}\right| - \mathfrak{K}\, \partial_r\ln \left|\frac{g_{tt}}{g_{\theta\theta}}\right| \right) \frac{dx}{\dot x}  \nonumber\\
&\propto \sum_k  \mathfrak{s}^{2k} 4^{-k} C^{2k}_k \left( \left(1-\mathfrak{s}^2\right) \, _2F_1\left(1,k+\frac{1}{2};k+1;\mathfrak{s}^2 \right) \right.  \nonumber\\
&\times \left. \partial_r(A_{2k}-D_{2k}) + \frac{\mathfrak{s}^2}{2k+2} \, _2F_1\left(1,k+\frac{1}{2};k+2;\mathfrak{s}^2 \right) \right.  \nonumber\\
&\times  \partial_r (A_{2k}-C_{2k}) \bigg{)}
+\partial_r \left( f r^{-2}\right)  \,,
\end{align}
with $\mathfrak{s} \equiv \sin\delta$. Comparing the equation above with Eq.~\eqref{eq:potential_m_generic}, we notice that the equations derived from two different approaches coincide if we identify $\mathfrak{s}^2 = 1-\alpha^2$. Thus, $\cos^{-1}(m/l)$ corresponds to the inclination angle of the orbit, and the potential peak to the averaged photon ring radius. 

Having formulated the correspondence between the peak of the eikonal effective potential and the averaged photon ring radius, we can use the results of sec.~\ref{subsec.sss} i.e., Eq.~\eqref{eikonalfrelya}, to formally build the eikonal correspondence between eikonal QNMs and the \textit{deformed} photon sphere in the deformed Schwarzschild spacetime.

\section{Examples}\label{sec.example}
In the previous sections, we have identified the correspondence between eikonal QNMs and trapped photon orbits in a deformed Schwarzschild spacetime \eqref{parametric}. The correspondence can be identified through the definiton of averaged photon ring radius. In this section, we will provide some simple examples to support our general results.

\subsection{Spherically symmetric deformations}

The first example for demonstrations is the eikonal correspondence in a Schwarzschild spacetime with spherically symmetric deformations. Such deformations belong to the axisymmetric deformations of Eq.~\eqref{parametric}, with $A_0(r)$ and $B_0(r)$ the only non-vanishing deformation functions. In this case, the expressions of the eikonal effective potential, i.e, Eqs.~\eqref{potentialll}, \eqref{potentialm0}, and \eqref{eq:potential_m_generic}, all reduce to
\begin{equation}
    V_{\textrm{eff}}(r)\approx l^2\frac{f}{r^2}\left(1+\epsilon A_0\right)\,.\label{eqssd}
\end{equation}
The peak of the effective potential is uniquely defined. For the trapped photon orbits, the spherical symmetry of the spacetime implies that the trapped photon orbits are always circular with radius determined by $\partial_r(g_{tt}/r^2)=0$. One can see that the peak of the effective potential \eqref{eqssd} is precisely at the trapped photon orbits. In fact, this correspondence can be identified in a similar manner for general spherically symmetric spacetimes, as we have already mentioned in sec.~\ref{sec.geoapp}.

\subsection{Supertranslated black hole}
The second set of examples we consider are two stationary axisymmetric black hole metrics that are diffeomorphic to the Schwarzschild black hole, commonly referred as soft-hair black holes.
Given the coordinate transformations, it is rather straightforward to derive the averaged photon ring radius and to test the correspondence by verifying the peak of the QNM potential.

\subsubsection{Comp\`ere-Long-Iofa type}
The first set of coordinate transformations is proposed by Comp\`ere and Long, later modified by Iofa (CLI) to focus on axisymmetric cases \cite{Compere:2016hzt,Iofa:2018pnf}. The CLI transformation is particularly interesting since only $\theta$ is transformed
\begin{align}
\sin\theta_S =  \sin\theta \sqrt{1-\left(\frac{C(\theta)}{K(r)}\right)^2} -\frac{C(\theta)}{K(r)} \cos\theta  \,,  \label{eq:IofaTrans}
\end{align}
where $\theta_S$ denotes the polar angle of the original Schwarzschild coordinates, $K(r)$ satisfies $dK(r)/dr=K(r)/[r\sqrt{f(r)}]$ \cite{Iofa:2018pnf}, $C(\theta) \equiv \sum_k \tilde{C}_k \cos^{2k} \theta \equiv \sum_k \tilde{C}_k x^{2k}$ is the deformation function, and $\tilde{C}_k$ are constant coefficients in the expansion. We neglect odd-parity deformations as they do not contribute to the QNM potential. Naively, under CLI transformation, only QNM waveforms are deformed. The potential should remain the same up to Darboux transformation, allowing us to put the projection method to test. 

Expanding the metric up to the first order of $C(\theta)$, we can identify the deformation functions used in Eq.~\eqref{parametric} through the relations \cite{Iofa:2018pnf}
\begin{align}
    A_j(r)&=B_j(r)=a_j(r)=b_j(r)=d_j(r)=e_j(r)=0\,,\nonumber\\
    C_j(r)&=\frac{8}{K(r)} \sum_k k \tilde{C}_k \left[ 2k \delta_j^{2k} - (2k-1) \delta_j^{2k-2} \right]\,,\nonumber\\
    D_j(r)&=\frac{8}{K(r)} \sum_k k \tilde{C}_k \delta_j^{2k}\,,\nonumber\\
    c_j(r)&=-\frac{4r}{\sqrt{f(r)}K(r)} \sum_k \frac{k\sqrt{1-x^2}}{x} \tilde{C}_k \delta_j^{2k}\,,
\end{align}
where $\delta$ is the Kronecker delta and the factor $k x^{-1} \sqrt{1-x^2}$ of $c_j$  transforms the parameter $e^j_{lm}$ to $d^j_{lm}$.

Plugging these relations into the effective potential \eqref{eq:Veff}, we get
\begin{align}
V_{\textrm{eff}}(r) &= l(l+1)\frac{f(r)}{r^2}+\frac{f(r)}{r}\frac{df(r)}{dr} +\epsilon\frac{2f(r)}{r^2}  \nonumber\\
&\times \sum_k \tilde{C}_k \left[ \frac{4 k }{K} M_1(k) +\partial_r \left(\frac{r\sqrt{f}}{K}\right) M_2(k) \right]  \,,
\end{align}
where
\begin{align}
M_1(k)\equiv&\, -a_{lm}^{2k} + 2k c_{lm}^{2k} - (2k-1) c_{lm}^{2k-2}  \nonumber\\
&+ \frac{2k-1}{2} d_{lm}^{2k} -\frac{2k-1}{2} d_{lm}^{2k-2}  \,,  \nonumber\\
M_2(k)\equiv&\, d_{lm}^{2k} -k(2k+1) b_{lm}^{2k} +k(2k-1) b_{lm}^{2k-2}\,.
\end{align}
Using integration by parts, we find $M_2(k)=0$. We also find $M_1(k)=0$ for all sets of $(k,l,m)$. Therefore, the effective QNM potential of CLI metric does not receive corrections from soft hairs up to the first order of $C(x)$. The QNM results reduce to Schwarzschild ones, and the peak of the effective potential in the eikonal limit is at $r=3M$. As for the trapped photon orbits, since CLI transformation acts on $\theta$ alone, the photon sphere or any sphere of a constant radius would remain a sphere of the same radius \cite{Lin:2022ksb}. As a result, the correspondence is clearly identified.

\subsubsection{Bondi-Metzner-Sachs type}
The previous example serves as a proving ground for QNM. Here we test the other side of the relation by considering another transformation proposed by Bondi, van der Burg, Metzner, and Sachs (BMS) \cite{Bondi:1962px,Sachs:1962wk,Sachs:1962zza},
\begin{align}
\delta v = C(x)  \,,\,\,  \delta r &= x \partial_xC(x) - (1-x^2) \partial_{xx}C(x)  \,,  \nonumber\\
\delta\theta &= - r^{-1} \sqrt{1-x^2} \partial_xC(x)  \,,  \label{eq:BMS_transform}
\end{align}
where $v = t + r^*$ is the advanced time \footnote{The deformation to the tortoise coordinate can be neglected as its effect on QNM potential is $O(\epsilon^2)$.}. For our purposes we have imposed the axisymmetry. Following the protocol, the deformation functions for $C(x) \equiv \sum_k \tilde{C}_k x^{2k}$ are
\begin{align}
A_j(r) &= \frac{f'(r)}{f(r)} \sum_k k \tilde{C}_k \left[ (2k+1) \delta_j^{2k} - (2k-1) \delta_j^{2k-2} \right]  \,,  \nonumber\\
C_j(r) &= -D_j(r) = \frac{1}{r} \sum_k 2k(2k-1) \tilde{C}_k \left(\delta_j^{2k-2}-\delta_j^{2k}\right)  \,.
\end{align}
To be concise, we only show relevant functions. Plugging them into Eq.~\eqref{eq:potential_m_generic}, the deformation to the effective potential of eikonal QNMs is
\begin{align}
\delta V_{\textrm{eff}} &= \frac{l^2}{r^4} \sum_k \frac{k C^{2k}_k \left( 1-\alpha ^2 \right)^k}{2^{2k-1}} \tilde{C}_k  \nonumber\\
&\times \Bigg{[} \frac{ 4 (r-2M) }{1-\alpha ^2} \alpha ^2 \,_2F_1 \left( 1, k-\frac{1}{2}; k; 1-\alpha^2\right)  \nonumber\\
&- 2(2k-1) (r-2M) \alpha ^2 \,_2F_1 \left( 1, k+\frac{1}{2}; k+1; 1-\alpha^2\right)  \nonumber\\
&+ (2k-1) (r-M) + 2M - \frac{ 2 (r-M) }{1-\alpha ^2} \Bigg{]}  \,.
\end{align}
For the first few $k$, the location of the potential peak will be shifted by $\frac{\tilde{C}_1}{2} (3\alpha^2-1)$, $\frac{3\tilde{C}_2}{4} (1-\alpha^2)(5\alpha^2-1) $, $\frac{15\tilde{C}_3}{16} \left(1-\alpha^2\right)^2 (7\alpha^2-1)$, $\frac{35\tilde{C}_4}{32} \left(1-\alpha^2\right)^3 (9\alpha^2-1)$, etc.

We now have one side of the correspondence. Let us derive the other side directly through the coordinate transformation. After applying Eq.~\eqref{eq:BMS_transform}, the photon sphere is slightly shifted to $r = 3M +  x \partial_xC(x) - (1-x^2) \partial_{xx}C(x)$. After fixing $C = x^{2k}$ and averaging over the background trajectory the averaged photon ring radius becomes \footnote{As argued previously, the trajectory deformation itself enters at $O(\epsilon^2)$.}
\begin{align}
3M +\frac{\mathfrak{s}^{2k} \left( 2k \left( 1-\mathfrak{s}^2 \right) - \mathfrak{s}^2 \right) \Gamma\left(k+\frac{1}{2}\right) }{ \sqrt{\pi} \mathfrak{s}^2 \Gamma(k)}  \,,
\end{align}
where $\Gamma$ is the Euler Gamma function and $\arcsin\mathfrak{s}$ is the inclination angle of the photon ring. One may verify that this formula indeed generates the sequence in the last paragraph after identifying $\mathfrak{s}^2$ as $1-\alpha^2$.

\section{Conclusions}\label{sec.con}

Based on the geometric optics approximation in generic curved spacetimes, the eikonal black hole QNMs are tightly related to the spherical photon orbits around the black hole. The violation of this eikonal correspondence can be a smoking gun of physics beyond GR. It is thus timely to understand how the eikonal correspondence could be broken in different circumstances.

It should be emphasized that the explicit identification of the eikonal correspondence for non-rotating black holes and rotating black holes (Kerr and Kerr-Newman spacetimes) in the literature substantially relies on the symmetries of the spacetime. One symmetry gives rise to the separability of master wave equations, and another gives rise to the separability of geodesic equations. However, there is no theoretical evidence that these symmetries are still preserved when going beyond GR. Therefore, to better understand how to utilize the eikonal correspondence for testing black hole models and gravitational theories, a thorough understanding of how to identify the correspondence in black hole spacetimes without sufficient symmetries is required.

In this work, we identify the well-defined eikonal correspondence for a Schwarzschild spacetime with a generic axisymmetric stationary deformation, neglecting frame-dragging effects. Although neither the master wave equations nor the geodesic equations are separable, the assumption that the spacetime deformations are small, i.e., only taking the $\epsilon$-order contributions into account, allows us to build the correspondence. Specifically, with this assumption, the radial part of the QNM master equation can be decoupled from the angular part, and recast into a Schr\"odinger-like form. The effective potential \eqref{eq:Veff} in the equation has a well-defined peak, which differs slightly from $r=3M$ due to the spacetime deformations. On the other hand, we find that the trapped photon orbits in the deformed Schwarzschild spacetime do not have a constant $r$ in general, even though circular orbits still exist and are parallel to the equatorial plane. The periodicity of the trapped photon orbits allows us to define the averaged radius of the orbits along one period. It turns out that in the eikonal limit, the peak radius of the effective QNM potential is identical to the averaged radius of the trapped photon orbits. The conclusion is valid for orbits with arbitrary inclinations.

The spacetime deformations we consider in this paper, although already very general, do not break the axisymmetry of the spacetime. It will be interesting to see whether a similar sense of eikonal correspondence can be identified or not when axisymmetry is broken. In addition, the assumption of the Schwarzschild spacetime as the reference for the order analysis is just for simplicity. Our ultimate goal is to understand the eikonal correspondence of a non-Kerr spacetime, with either small or moderate deformations from the Kerr metric. Similar to non-spinning black hole spacetimes, Kerr spacetimes have separable geodesic equations. Their spherical photon orbits do have constant radii when expressed in some proper coordinates. Therefore, our results that the eikonal correspondence can be identified through the definition of the averaged radius of trapped photon orbits in deformed Schwarzschild spacetimes may also give some hints when considering deformed Kerr spacetimes. The eikonal correspondence can also be helpful in the gravitational waveform modelings of black hole mergers \cite{McWilliams:2018ztb}. We hope to address these issues elsewhere in the future.

\acknowledgements
The authors would like to thank Feng-Li Lin and Avani Patel for the fruitful discussions during the development of this work. CYC is supported by the Institute of Physics of
Academia Sinica. HWC is supported by Ministry of Science and Technology (MoST) of Taiwan grant number 111-2811-M-002-048 through department of physics and the Leung Center for Cosmology and Particle Astrophysics (LeCosPA) of National Taiwan University. JST is supported by MoST through grant number 109-2112-M-003-007-MY3.


\begin{thebibliography}{99} 

\bibitem{LIGOScientific:2016aoc}
B.~P.~Abbott \textit{et al.} [LIGO Scientific and Virgo],
Phys. Rev. Lett. \textbf{116}, no.6, 061102 (2016).


\bibitem{LIGOScientific:2018mvr}
B.~P.~Abbott \textit{et al.} [LIGO Scientific and Virgo],
Phys. Rev. X \textbf{9}, no.3, 031040 (2019).


\bibitem{LIGOScientific:2020ibl}
R.~Abbott \textit{et al.} [LIGO Scientific and Virgo],
Phys. Rev. X \textbf{11}, 021053 (2021).


\bibitem{LIGOScientific:2021djp}
R.~Abbott \textit{et al.} [LIGO Scientific, VIRGO and KAGRA],
[arXiv:2111.03606 [gr-qc]].

\bibitem{Barack:2018yly}
L.~Barack, V.~Cardoso, S.~Nissanke, T.~P.~Sotiriou, A.~Askar, C.~Belczynski, G.~Bertone, E.~Bon, D.~Blas and R.~Brito, \textit{et al.}
Class. Quant. Grav. \textbf{36}, no.14, 143001 (2019).


\bibitem{Berti:2018vdi}
E.~Berti, K.~Yagi, H.~Yang and N.~Yunes,
Gen. Rel. Grav. \textbf{50}, no.5, 49 (2018).


\bibitem{Kokkotas:1999bd}
K.~D.~Kokkotas and B.~G.~Schmidt,
Living Rev. Rel. \textbf{2}, 2 (1999).


\bibitem{Berti:2009kk}
E.~Berti, V.~Cardoso and A.~O.~Starinets,
Class. Quant. Grav. \textbf{26}, 163001 (2009).

\bibitem{Konoplya:2011qq}
R.~A.~Konoplya and A.~Zhidenko,
Rev. Mod. Phys. \textbf{83}, 793-836 (2011).







\bibitem{Dreyer:2003bv}
O.~Dreyer, B.~J.~Kelly, B.~Krishnan, L.~S.~Finn, D.~Garrison and R.~Lopez-Aleman,
Class. Quant. Grav. \textbf{21}, 787-804 (2004).

\bibitem{Blazquez-Salcedo:2017txk}
J.~L.~Bl\'azquez-Salcedo, F.~S.~Khoo and J.~Kunz,
Phys. Rev. D \textbf{96}, no.6, 064008 (2017).

\bibitem{Glampedakis:2017dvb}
K.~Glampedakis, G.~Pappas, H.~O.~Silva and E.~Berti,
Phys. Rev. D \textbf{96}, no.6, 064054 (2017).

\bibitem{Tattersall:2018nve}
O.~J.~Tattersall and P.~G.~Ferreira,
Phys. Rev. D \textbf{97}, no.10, 104047 (2018).

\bibitem{Cardoso:2019mqo}
V.~Cardoso, M.~Kimura, A.~Maselli, E.~Berti, C.~F.~B.~Macedo and R.~McManus,
Phys. Rev. D \textbf{99}, no.10, 104077 (2019).

\bibitem{Chen:2019iuo}
C.~Y.~Chen and P.~Chen,
Phys. Rev. D \textbf{99}, no.10, 104003 (2019).

\bibitem{Isi:2019aib}
M.~Isi, M.~Giesler, W.~M.~Farr, M.~A.~Scheel and S.~A.~Teukolsky,
Phys. Rev. Lett. \textbf{123}, no.11, 111102 (2019).

\bibitem{McManus:2019ulj}
R.~McManus, E.~Berti, C.~F.~B.~Macedo, M.~Kimura, A.~Maselli and V.~Cardoso,
Phys. Rev. D \textbf{100}, no.4, 044061 (2019).

\bibitem{Maselli:2019mjd}
A.~Maselli, P.~Pani, L.~Gualtieri and E.~Berti,
Phys. Rev. D \textbf{101}, no.2, 024043 (2020).

\bibitem{Cabero:2019zyt}
M.~Cabero, J.~Westerweck, C.~D.~Capano, S.~Kumar, A.~B.~Nielsen and B.~Krishnan,
Phys. Rev. D \textbf{101}, no.6, 064044 (2020).

\bibitem{Tattersall:2019nmh}
O.~J.~Tattersall,
Class. Quant. Grav. \textbf{37}, no.11, 115007 (2020).

\bibitem{Bouhmadi-Lopez:2020oia}
M.~Bouhmadi-L\'opez, S.~Brahma, C.~Y.~Chen, P.~Chen and D.~h.~Yeom,
JCAP \textbf{07}, 066 (2020).

\bibitem{Chen:2020evr}
C.~Y.~Chen, Y.~H.~Kung and P.~Chen,
Phys. Rev. D \textbf{102}, no.12, 124033 (2020).

\bibitem{Isi:2020tac}
M.~Isi, W.~M.~Farr, M.~Giesler, M.~A.~Scheel and S.~A.~Teukolsky,
Phys. Rev. Lett. \textbf{127}, no.1, 011103 (2021).

\bibitem{Arbey:2021jif}
A.~Arbey, J.~Auffinger, M.~Geiller, E.~R.~Livine and F.~Sartini,
Phys. Rev. D \textbf{103}, no.10, 104010 (2021).

\bibitem{Bamber:2021knr}
J.~Bamber, O.~J.~Tattersall, K.~Clough and P.~G.~Ferreira,
Phys. Rev. D \textbf{103}, no.12, 124013 (2021).

\bibitem{Chen:2021cts}
C.~Y.~Chen, M.~Bouhmadi-L\'opez and P.~Chen,
Eur. Phys. J. Plus \textbf{136}, no.2, 253 (2021).

\bibitem{Pierini:2021jxd}
L.~Pierini and L.~Gualtieri,
Phys. Rev. D \textbf{103}, 124017 (2021).

\bibitem{Ikeda:2021uvc}
T.~Ikeda, M.~Bianchi, D.~Consoli, A.~Grillo, J.~F.~Morales, P.~Pani and G.~Raposo,
Phys. Rev. D \textbf{104}, no.6, 066021 (2021).

\bibitem{Ghosh:2021mrv}
A.~Ghosh, R.~Brito and A.~Buonanno,
Phys. Rev. D \textbf{103}, no.12, 124041 (2021).

\bibitem{Cano:2021myl}
P.~A.~Cano, K.~Fransen, T.~Hertog and S.~Maenaut,
Phys. Rev. D \textbf{105}, no.2, 024064 (2022).

\bibitem{Momennia:2022tug}
M.~Momennia,
[arXiv:2204.03259 [gr-qc]].

\bibitem{LIGOScientific:2020tif}
R.~Abbott \textit{et al.} [LIGO Scientific and Virgo],
Phys. Rev. D \textbf{103}, no.12, 122002 (2021).

\bibitem{LIGOScientific:2021sio}
R.~Abbott \textit{et al.} [LIGO Scientific, VIRGO and KAGRA],
[arXiv:2112.06861 [gr-qc]].


\bibitem{EventHorizonTelescope:2019dse}
K.~Akiyama \textit{et al.} [Event Horizon Telescope],
Astrophys. J. Lett. \textbf{875}, L1 (2019).

\bibitem{Cunha:2018acu}
P.~V.~P.~Cunha and C.~A.~R.~Herdeiro,
Gen. Rel. Grav. \textbf{50}, no.4, 42 (2018).

\bibitem{Perlick:2021aok}
V.~Perlick and O.~Y.~Tsupko,
Phys. Rept. \textbf{947}, 1-39 (2022).




\bibitem{Johannsen:2013vgc}
T.~Johannsen,
Astrophys. J. \textbf{777}, 170 (2013).

\bibitem{Li:2013jra}
Z.~Li and C.~Bambi,
JCAP \textbf{01}, 041 (2014).

\bibitem{Younsi:2016azx}
Z.~Younsi, A.~Zhidenko, L.~Rezzolla, R.~Konoplya and Y.~Mizuno,
Phys. Rev. D \textbf{94}, no.8, 084025 (2016).

\bibitem{Wang:2017hjl}
M.~Wang, S.~Chen and J.~Jing,
JCAP \textbf{10}, 051 (2017).


\bibitem{Tsukamoto:2017fxq}
N.~Tsukamoto,
Phys. Rev. D \textbf{97}, no.6, 064021 (2018).

\bibitem{Abdikamalov:2019ztb}
A.~B.~Abdikamalov, A.~A.~Abdujabbarov, D.~Ayzenberg, D.~Malafarina, C.~Bambi and B.~Ahmedov,
Phys. Rev. D \textbf{100}, no.2, 024014 (2019).

\bibitem{Shaikh:2019fpu}
R.~Shaikh,
Phys. Rev. D \textbf{100}, no.2, 024028 (2019).


\bibitem{Bambi:2019tjh}
C.~Bambi, K.~Freese, S.~Vagnozzi and L.~Visinelli,
Phys. Rev. D \textbf{100}, no.4, 044057 (2019).

\bibitem{Vagnozzi:2019apd}
S.~Vagnozzi and L.~Visinelli,
Phys. Rev. D \textbf{100}, no.2, 024020 (2019).

\bibitem{Kumar:2019pjp}
R.~Kumar, S.~G.~Ghosh and A.~Wang,
Phys. Rev. D \textbf{100}, no.12, 124024 (2019).

\bibitem{Liu:2020ola}
C.~Liu, T.~Zhu, Q.~Wu, K.~Jusufi, M.~Jamil, M.~Azreg-A\"\i{}nou and A.~Wang,
Phys. Rev. D \textbf{101}, no.8, 084001 (2020)
[erratum: Phys. Rev. D \textbf{103}, no.8, 089902 (2021)].


\bibitem{EslamPanah:2020hoj}
B.~Eslam Panah, K.~Jafarzade and S.~H.~Hendi,
Nucl. Phys. B \textbf{961}, 115269 (2020).

\bibitem{Khodadi:2020jij}
M.~Khodadi, A.~Allahyari, S.~Vagnozzi and D.~F.~Mota,
JCAP \textbf{09}, 026 (2020).

\bibitem{Jusufi:2020odz}
K.~Jusufi, M.~Azreg-A\"\i{}nou, M.~Jamil, S.~W.~Wei, Q.~Wu and A.~Wang,
Phys. Rev. D \textbf{103}, no.2, 024013 (2021).

\bibitem{EventHorizonTelescope:2020qrl}
D.~Psaltis \textit{et al.} [Event Horizon Telescope],
Phys. Rev. Lett. \textbf{125}, no.14, 141104 (2020).

\bibitem{Khodadi:2020gns}
M.~Khodadi and E.~N.~Saridakis,
Phys. Dark Univ. \textbf{32}, 100835 (2021).

\bibitem{Hu:2020usx}
Z.~Hu, Z.~Zhong, P.~C.~Li, M.~Guo and B.~Chen,
Phys. Rev. D \textbf{103}, no.4, 044057 (2021).



\bibitem{Brahma:2020eos}
S.~Brahma, C.~Y.~Chen and D.~h.~Yeom,
Phys. Rev. Lett. \textbf{126}, no.18, 181301 (2021).


\bibitem{Lima:2021las}
H.~C.~D.~Lima, Junior., L.~C.~B.~Crispino, P.~V.~P.~Cunha and C.~A.~R.~Herdeiro,
Phys. Rev. D \textbf{103}, no.8, 084040 (2021).

\bibitem{Konoplya:2021slg}
R.~A.~Konoplya and A.~Zhidenko,
Phys. Rev. D \textbf{103}, no.10, 104033 (2021).

\bibitem{EventHorizonTelescope:2021dqv}
P.~Kocherlakota \textit{et al.} [Event Horizon Telescope],
Phys. Rev. D \textbf{103}, no.10, 104047 (2021).

\bibitem{Lara:2021zth}
G.~Lara, S.~H.~V\"olkel and E.~Barausse,
Phys. Rev. D \textbf{104}, no.12, 124041 (2021).

\bibitem{Cimdiker:2021cpz}
\.I.~\c{C}imdiker, D.~Demir and A.~\"Ovg\"un,
Phys. Dark Univ. \textbf{34}, 100900 (2021).

\bibitem{Meng:2022kjs}
Y.~Meng, X.~M.~Kuang and Z.~Y.~Tang,
[arXiv:2204.00897 [gr-qc]].


\bibitem{Chen:2020aix}
C.~Y.~Chen,
JCAP \textbf{05}, 040 (2020).

\bibitem{Eichhorn:2021etc}
A.~Eichhorn and A.~Held,
Eur. Phys. J. C \textbf{81}, no.10, 933 (2021).

\bibitem{Eichhorn:2021iwq}
A.~Eichhorn and A.~Held,
JCAP \textbf{05}, 073 (2021).



\bibitem{Lin:2022ksb}
F.~L.~Lin, A.~Patel and H.~Y.~Pu,
[arXiv:2202.13559 [gr-qc]].


\bibitem{Eichhorn:2022oma}
A.~Eichhorn, A.~Held and P.~V.~Johannsen,
[arXiv:2204.02429 [gr-qc]].














\bibitem{Ferrari:1984zz}
V.~Ferrari and B.~Mashhoon,
Phys. Rev. D \textbf{30}, 295-304 (1984).


\bibitem{Hod:2009td}
S.~Hod,
Phys. Rev. D \textbf{80}, 064004 (2009).


\bibitem{Gaddam:2020mwe}
N.~Gaddam and N.~Groenenboom,
[arXiv:2012.02355 [hep-th]].


\bibitem{Cardoso:2008bp}
V.~Cardoso, A.~S.~Miranda, E.~Berti, H.~Witek and V.~T.~Zanchin,
Phys. Rev. D \textbf{79}, no.6, 064016 (2009).


\bibitem{Dolan:2009nk}
S.~R.~Dolan and A.~C.~Ottewill,
Class. Quant. Grav. \textbf{26}, 225003 (2009).


\bibitem{Dolan:2010wr}
S.~R.~Dolan,
Phys. Rev. D \textbf{82}, 104003 (2010).

\bibitem{Yang:2012he}
H.~Yang, D.~A.~Nichols, F.~Zhang, A.~Zimmerman, Z.~Zhang and Y.~Chen,
Phys. Rev. D \textbf{86}, 104006 (2012).

\bibitem{Li:2021zct}
P.~C.~Li, T.~C.~Lee, M.~Guo and B.~Chen,
Phys. Rev. D \textbf{104}, no.8, 084044 (2021).






\bibitem{Stefanov:2010xz}
I.~Z.~Stefanov, S.~S.~Yazadjiev and G.~G.~Gyulchev,
Phys. Rev. Lett. \textbf{104}, 251103 (2010).




\bibitem{Jusufi:2019ltj}
K.~Jusufi,
Phys. Rev. D \textbf{101}, no.8, 084055 (2020).

\bibitem{Jusufi:2020dhz}
K.~Jusufi,
Phys. Rev. D \textbf{101}, no.12, 124063 (2020).

\bibitem{Cuadros-Melgar:2020kqn}
B.~Cuadros-Melgar, R.~D.~B.~Fontana and J.~de Oliveira,
Phys. Lett. B \textbf{811}, 135966 (2020).


\bibitem{Yang:2021zqy}
H.~Yang,
Phys. Rev. D \textbf{103}, no.8, 084010 (2021).





\bibitem{Li:2021mnx}
S.~Li, A.~A.~Abdujabbarov and W.~B.~Han,
Eur. Phys. J. C \textbf{81}, no.7, 649 (2021).


\bibitem{Zhang:2021ygh}
Z.~Zhang,
Class. Quant. Grav. \textbf{39}, no.1, 015003 (2022).


\bibitem{Konoplya:2017wot}
R.~A.~Konoplya and Z.~Stuchl\'\i{}k,
Phys. Lett. B \textbf{771}, 597-602 (2017).


\bibitem{Glampedakis:2019dqh}
K.~Glampedakis and H.~O.~Silva,
Phys. Rev. D \textbf{100}, no.4, 044040 (2019).

\bibitem{Chen:2019dip}
C.~Y.~Chen and P.~Chen,
Phys. Rev. D \textbf{101}, no.6, 064021 (2020).

\bibitem{Silva:2019scu}
H.~O.~Silva and K.~Glampedakis,
Phys. Rev. D \textbf{101}, no.4, 044051 (2020).


\bibitem{Moura:2021eln}
F.~Moura and J.~Rodrigues,
Phys. Lett. B \textbf{819}, 136407 (2021).


\bibitem{Bryant:2021xdh}
A.~Bryant, H.~O.~Silva, K.~Yagi and K.~Glampedakis,
Phys. Rev. D \textbf{104}, no.4, 044051 (2021).







\bibitem{Cano:2020cao}
P.~A.~Cano, K.~Fransen and T.~Hertog,
Phys. Rev. D \textbf{102}, no.4, 044047 (2020).


\bibitem{Glampedakis:2018blj}
K.~Glampedakis and G.~Pappas,
Phys. Rev. D \textbf{99}, no.12, 124041 (2019).












\bibitem{MTW:Gravitation}
C.~W.~Misner, K.~S.~Thorne and J.~A.~Wheeler, \textit{Gravitation}, (W. H. Freeman and Company, New York, 1973).





\bibitem{Mashhoon:1982im}
B.~Mashhoon,
``QUASINORMAL MODES OF A BLACK HOLE,'' (1982).

\bibitem{Schutz:1985km}
B.~F.~Schutz and C.~M.~Will,
Astrophys. J. Lett. \textbf{291}, L33-L36 (1985).

\bibitem{Iyer:1986np}
S.~Iyer and C.~M.~Will,
Phys. Rev. D \textbf{35}, 3621 (1987).










\bibitem{Anson:2020trg}
T.~Anson, E.~Babichev, C.~Charmousis and M.~Hassaine,
JHEP \textbf{01}, 018 (2021).


\bibitem{BenAchour:2020fgy}
J.~Ben Achour, H.~Liu, H.~Motohashi, S.~Mukohyama and K.~Noui,
JCAP \textbf{11}, 001 (2020).

\bibitem{Minamitsuji:2020jvf}
M.~Minamitsuji,
Phys. Rev. D \textbf{102}, no.12, 124017 (2020).


\bibitem{Delaporte:2022acp}
H.~Delaporte, A.~Eichhorn and A.~Held,
[arXiv:2203.00105 [gr-qc]].










\bibitem{Frolov:1998wf}
V.~P.~Frolov and I.~D.~Novikov,
``Black hole physics: Basic concepts and new developments,''
doi:10.1007/978-94-011-5139-9


\bibitem{Cardoso:2021qqu}
V.~Cardoso and A.~Foschi,
Phys. Rev. D \textbf{104}, no.2, 024004 (2021).


\bibitem{Antonov1981}
V.~A.~Antonov and K.~V.~Hol\v{s}evnikov, 
``An estimate of the remainder in the expansion of the generating function for the Legendre polynomials (generalization and improvement of Bernstein's inequality),"
Vestnik Leningrad. Univ. Mat.13(1981), 163-166.

\bibitem{Lorch1993}
L.~Lorch,  
``Alternative proof of a sharpened form of Bernstein's inequality for Legendre polynomials,"
Appl. Anal. 14 (1983), 237-240.

\bibitem{Antonov:2010}
V.~A.~Antonov, K.~V.~Kholshevnikov and V.~S.~Shaidulin, 
``Estimating the derivative of the Legendre polynomial," Vestnik St.Petersb. Univ.Math. \textbf{43}, 191-197 (2010).



\bibitem[Ragni et al.(2010)]{2010IJQC..110..731R} Ragni, M., Bitencourt, A.~C.~P., da S. Ferreira, C., et al.\ 2010, International Journal of Quantum Chemistry, 110, 731. 







\bibitem{Compere:2016hzt}
G.~Comp\`ere and J.~Long,
Class. Quant. Grav. \textbf{33}, no.19, 195001 (2016).




\bibitem{Iofa:2018pnf}
M.~Z.~Iofa,
Phys. Rev. D \textbf{99}, no.6, 064052 (2019).




\bibitem{Bondi:1962px}
H.~Bondi, M.~G.~J.~van der Burg and A.~W.~K.~Metzner,
Proc. Roy. Soc. Lond. A \textbf{269}, 21-52 (1962).


\bibitem{Sachs:1962wk}
R.~K.~Sachs,
Proc. Roy. Soc. Lond. A \textbf{270}, 103-126 (1962).


\bibitem{Sachs:1962zza}
R.~Sachs,
Phys. Rev. \textbf{128}, 2851-2864 (1962).






\bibitem{McWilliams:2018ztb}
S.~T.~McWilliams,
Phys. Rev. Lett. \textbf{122}, no.19, 191102 (2019).


\end{thebibliography}
\end{document}